\documentclass[twocolumn,10pt]{article}
\usepackage{xcolor}
\usepackage[noblocks]{authblk}
\usepackage{tabularx}
\usepackage{graphicx}
\usepackage{dcolumn}
\usepackage{bm}
\usepackage{amsmath}
\usepackage{amssymb}
\usepackage[switch]{lineno} 
\usepackage[squaren,Gray]{SIunits}
\usepackage{soul}
\usepackage{hyperref}
\usepackage{array}
\newcolumntype{L}[1]{>{\raggedright\let\newline\\\arraybackslash\hspace{0pt}}m{#1}}
\newcolumntype{C}[1]{>{\centering\let\newline\\\arraybackslash\hspace{0pt}}m{#1}}
\newcolumntype{R}[1]{>{\raggedleft\let\newline\\\arraybackslash\hspace{0pt}}m{#1}}
\let\OLDthebibliography\thebibliography
\renewcommand\thebibliography[1]{
  \OLDthebibliography{#1}
  \setlength{\parskip}{0pt}
  \setlength{\itemsep}{0pt plus 0.3ex}
}
\makeatletter
\def\@maketitle{%
  \newpage
  \null
  \vskip 0.em%
  \begin{center}%
  \let \footnote \thanks
    {\huge \bfseries \@title \par}%
    \vskip .0em%
    {\normalsize
      \lineskip 5.em%
      \begin{tabular}[t]{c}%
        \@author
      \end{tabular}\par}%
    \vskip 0.em%
    {\normalsize \@date}%
  \end{center}%
  \par
  \vskip .em}
\makeatother
\newcommand{\ang}{$\theta _{13} $}
\newcommand{\sang}{$\rm{sin^2 2} \theta_{13} $}

\newcommand{\Aachen}{III. Physikalisches Institut, RWTH Aachen University, 52056 Aachen, Germany}
\newcommand{\Alabama}{Department of Physics and Astronomy, University of Alabama, Tuscaloosa, Alabama 35487, USA}
\newcommand{\Argonne}{Argonne National Laboratory, Argonne, Illinois 60439, USA}
\newcommand{\APC}{
APC, CNRS/IN2P3, CEA/IRFU, Observatoire de Paris, Sorbonne Paris Cit\'{e} University,
75205 Paris Cedex 13, France}
\newcommand{\CBPF}{Centro Brasileiro de Pesquisas F\'{i}sicas, Rio de Janeiro, RJ, 22290-180, Brazil}
\newcommand{\CENBG}{CENBG, CNRS/IN2P3, Universit\'e de Bordeaux, F-33175 Gradignan, France}
\newcommand{\Chicago}{The Enrico Fermi Institute, The University of Chicago, Chicago, Illinois 60637, USA}
\newcommand{\CIEMAT}{Centro de Investigaciones Energ\'{e}ticas, Medioambientales y Tecnol\'{o}gicas, CIEMAT, 28040, Madrid, Spain}

\newcommand{\Drexel}{Department of Physics, Drexel University, Philadelphia, Pennsylvania 19104, USA}

\newcommand{\INR}{Institute of Nuclear Research of the Russian Academy of Sciences, Moscow 117312, Russia}
\newcommand{\CEA}{IRFU, CEA, Universit\'{e} Paris-Saclay, 91191 Gif-sur-Yvette, France}

\newcommand{\Kitasato}{Department of Physics, Kitasato University, Sagamihara, 252-0373, Japan}
\newcommand{\Kobe}{Department of Physics, Kobe University, Kobe, 657-8501, Japan}
\newcommand{\Kurchatov}{NRC Kurchatov Institute, 123182 Moscow, Russia}
\newcommand{\MIT}{Massachusetts Institute of Technology, Cambridge, Massachusetts 02139, USA}
\newcommand{\MaxPlanck}{Max-Planck-Institut f\"{u}r Kernphysik, 69117 Heidelberg, Germany}
\newcommand{\NotreDame}{University of Notre Dame, Notre Dame, Indiana 46556, USA}
\newcommand{\IPHC}{IPHC, CNRS/IN2P3, Universit\'{e} de Strasbourg, 67037 Strasbourg, France}
\newcommand{\SUBATECH}{SUBATECH, CNRS/IN2P3, Universit\'{e} de Nantes, IMT-Atlantique, 44307 Nantes, France}

\newcommand{\TohokuUni}{Research Center for Neutrino Science, Tohoku University, Sendai 980-8578, Japan}

\newcommand{\TokyoInst}{Department of Physics, Tokyo Institute of Technology, Tokyo, 152-8551, Japan }
\newcommand{\TokyoMet}{Department of Physics, Tokyo Metropolitan University, Tokyo, 192-0397, Japan}
\newcommand{\Muenchen}{Physik Department, Technische Universit\"{a}t M\"{u}nchen, 85748 Garching, Germany}
\newcommand{\Tubingen}{Kepler Center for Astro and Particle Physics, Universit\"{a}t T\"{u}bingen, 72076 T\"{u}bingen, Germany}

\newcommand{\UNICAMP}{Universidade Estadual de Campinas-UNICAMP, Campinas, SP, 13083-970, Brazil}
\newcommand{\vtech}{Center for Neutrino Physics, Virginia Tech, Blacksburg, Virginia 24061, USA}
\newcommand{\Chooz}{{LNCA Underground Laboratory, IN2P3/CNRS - CEA, Chooz, France}}
\newcommand{\Londrina}{Universidade Estadual de Londrina, 86057-970 Londrina, Brazil}
\newcommand{\Hawaii}{Physics \& Astronomy Department, University of Hawaii at Manoa, Honolulu, Hawaii 96822, USA}
\newcommand{\IFIC}{Instituto de F\'{i}sica Corpuscular, IFIC (CSIC/UV), 46980 Paterna, Spain}
\newcommand{\KEK}{High Energy Accelerator Research Organization (KEK), Tsukuba, Ibaraki, Japan}
\newcommand{\StonyBrooks}{State University of New York at Stony Brook, Stony Brook, NY,  11755, USA}

\newcommand{\SK}{Kamioka Observatory, Institute for Cosmic Ray Research, University of Tokyo, Kamioka, Gifu 506-1205, Japan}

\newcommand{\SD}{South Dakota School of Mines \& Technology, 501 E. Saint Joseph St.~Rapid City, SD 57701}
\newcommand{\Arcadia}{Physics Department, Arcadia University, 450 S. Easton Road, Glenside, PA 19038}
\newcommand{\TokyoScience}{Tokyo University of Science, Noda, Chiba, Japan}
\newcommand{\LAPP}{Laboratoire d'Annecy-le-Vieux de physique des particules (LAPP), CNRS/IN2P3 , 74940 Annecy-le-Vieux, France}
\newcommand{\LAL}{Laboratoire de l'Acc\'{e}l\'{e}rateur Lin\'{e}aire (LAL), CNRS/IN2P3, F-91405 Orsay, France}
\newcommand{\Now}{
[1]~\LAL, 
[2]~\Londrina, 
[3]~\MIT,
[4]~\TokyoScience,
[5]~\Hawaii,
[6]~\KEK,
[7]~\Arcadia,
[8]~\LAPP,
[9]~\IFIC,
[10]~\SK,
[11]~\SD
~and 
[12]~\StonyBrooks.
\\
Contact:~\texttt{anatael@in2p3.fr \& christian.buck@mpi-hd.mpg.de}.
}
\begin{document}
%
%
%
%
%
\author[d]{H.\,de\,Kerret\thanks{Deceased.}} 
\author[e]{T.~Abrah\~{a}o}
\author[o]{H.~Almazan} 
\author[e]{J.\,C.~dos Anjos} 
\author[v]{S.~Appel} 
\author[k]{{J.\,C.~Barriere}} 
\author[a]{I.~Bekman} 
\author[r]{T.\,J.\,C.~Bezerra} 
\author[j]{L.~Bezrukov} 
\author[g]{E.~Blucher} 
\author[q]{T.~Brugi\`{e}re} 
\author[o]{C.~Buck} 
\author[b]{J.~Busenitz} 
\author[d,aa]{A.~Cabrera\thanks{\Now}$^{1,}$} 
\author[h]{M.~Cerrada} 
\author[f]{E.~Chauveau} 
\author[e,$\dagger$2]{P.~Chimenti}
\author[k]{{O.~Corpace}} 
\author[d]{J.\,V.~Dawson} 
\author[c]{Z.~Djurcic} 
\author[n]{A.~Etenko} 
\author[s]{H.~Furuta} 
\author[h]{I.~Gil-Botella} 
\author[d]{{A.~Givaudan}} 
\author[d]{{H.~Gomez}} 
\author[y]{L.\,F.\,G.~Gonzalez} 
\author[c]{M.\,C.~Goodman} 
\author[m]{T.~Hara} 
\author[o]{{J.~Haser}} 
\author[a]{D.~Hellwig} 
\author[d,$\dagger$3]{A.~Hourlier}
\author[t,$\dagger$4]{M.~Ishitsuka}
\author[w]{J.~Jochum} 
\author[f]{C.~Jollet} 
\author[f,q]{K.~Kale} 
\author[t]{M.~Kaneda} 
\author[d]{{M.~Karakac}} 
\author[l]{T.~Kawasaki} 
\author[y]{E.~Kemp} 
\author[d]{D.~Kryn} 
\author[t]{M.~Kuze} 
\author[w]{T.~Lachenmaier} 
\author[i]{C.\,E.~Lane} 
\author[k,d]{T.~Lasserre} 
\author[h]{C.~Lastoria} 
\author[k]{D.~Lhuillier} 
\author[e]{H.\,P.~Lima Jr} 
\author[o]{M.~Lindner} 
\author[h]{J.\,M.~L\'opez-Casta\~no} 
\author[p]{J.\,M.~LoSecco} 
\author[j]{B.~Lubsandorzhiev} 
\author[u,m]{J.~Maeda} 
\author[z]{C.~Mariani} 
\author[i,$\dagger$5]{J.~Maricic} 
\author[r]{J.~Martino} 
\author[u,$\dagger$6]{T.~Matsubara}
\author[k]{G.~Mention} 
\author[f]{A.~Meregaglia} 
\author[i,$\dagger$7]{T.~Miletic}
\author[i,$\dagger$5]{R.~Milincic} 
\author[k,$\dagger$8]{A.~Minotti}
\author[h]{D.~Navas-Nicol\'as} 
\author[h,$\dagger$9]{P.~Novella}
\author[v]{L.~Oberauer} 
\author[d]{M.~Obolensky} 
\author[d,k]{A.~Onillon} 
\author[n]{A.~Oralbaev} 
\author[h]{C.~Palomares} 
\author[e]{I.\,M.~Pepe} 
\author[r,$\dagger$10]{G.~Pronost}
\author[b,$\dagger$11]{J.~Reichenbacher}
\author[o,$\dagger$5]{B.~Reinhold} 
\author[v]{S.~Sch\"{o}nert} 
\author[o]{S.~Schoppmann} 
\author[k]{{L.~Scola}} 
\author[t]{R.~Sharankova} 
\author[k,$\dagger$3]{V.~Sibille} 
\author[j]{V.~Sinev} 
\author[n]{M.~Skorokhvatov} 
\author[a]{P.~Soldin} 
\author[a]{A.~Stahl} 
\author[b]{I.~Stancu} 
\author[w]{L.\,F.\,F.~Stokes} 
\author[s,d]{F.~Suekane} 
\author[n]{S.~Sukhotin} 
\author[u]{T.~Sumiyoshi} 
\author[b,4]{Y.~Sun} 
\author[k]{{C.~Veyssiere}} 
\author[r]{B.~Viaud} 
\author[k]{M.~Vivier} 
\author[d,e]{S.~Wagner} 
\author[a]{C.~Wiebusch} 
\author[c,$\dagger$12]{G.~Yang}
\author[r]{F.~Yermia} 
%
%
%
\affil[a]{\Aachen} 
\affil[b]{\Alabama} 
\affil[c]{\Argonne} 
\affil[d]{\APC} 
\affil[e]{\CBPF} 
\affil[f]{\CENBG} 
\affil[g]{\Chicago} 
\affil[h]{\CIEMAT} 
\affil[i]{\Drexel} 
\affil[j]{\INR} 
\affil[k]{\CEA} 
\affil[l]{\Kitasato} 
\affil[m]{\Kobe} 
\affil[n]{\Kurchatov} 
\affil[o]{\MaxPlanck} 
\affil[p]{\NotreDame} 
\affil[q]{\IPHC} 
\affil[r]{\SUBATECH} 
\affil[s]{\TohokuUni} 
\affil[t]{\TokyoInst} 
\affil[u]{\TokyoMet} 
\affil[v]{\Muenchen} 
\affil[w]{\Tubingen} 
\affil[y]{\UNICAMP} 
\affil[z]{\vtech} 
\affil[aa]{\Chooz} 
\affil[$\dagger$]{Current Institutions}
%
%
\title{\bf First Double Chooz $\mathbf{\theta_{13}}$ Measurement via Total Neutron Capture Detection\\
\vspace{0.3cm}
{\large The Double Chooz Collaboration}
}
\maketitle

{\bf
\noindent
Neutrinos were assumed to be massless particles until the discovery of  \emph{neutrino oscillation} process.
This phenomenon indicates that the neutrinos have non-zero masses and  the mass eigenstates ($\mathbf{\nu_1}$, $\mathbf{\nu_2}$, $\mathbf{\nu_3}$)  are mixing of their flavour eigenstates ($\mathbf{\nu_e}$, $\mathbf{\nu_\mu}$, $\mathbf{\nu_\tau}$). 
The oscillations between different flavour eigenstates are described by 
three mixing angles ($\mathbf{\theta_{12}}$, $\mathbf{\theta_{23}}$, $\mathbf{\theta_{13}}$),  two differences of the square of neutrino masses 
($\mathbf{\Delta m_{21}^2}$, $\mathbf{\Delta m_{31}^2}$) and a  charge conjugation parity symmetry (CP) violating phase $\delta_{\rm CP}$. 
The Double Chooz (DC) experiment, located near the Chooz Electricit\'e de France reactors, France, measures the oscillation parameter $\mathbf{\theta_{13}}$ using reactor neutrinos. 
In this paper, DC reports its latest $\mathbf{\theta_{13}}$ result, $\mathbf{\sin^2(2\theta_{13})=0.105\pm0.014}$, exploiting its multi-detector configuration, iso-flux baseline, reactor-off data and a novel total neutron capture detection technique. 
Since $\mathbf{\theta_{13}}$ was the last unknown mixing angle, which is necessary to measure $\delta_{\rm CP}$,  the result has contributed to the completion of a quest of the neutrino oscillation studies lasting half a century and to pave way toward the CP violation measurement.
In addition, DC provides the most precise measurement of the reactor neutrino flux 
to date, given by the mean cross section per fission $\mathbf{\langle \sigma_f\rangle = (5.71\pm0.06)\times10^{-43}}$cm$^{\mathbf2}$/fission.}
\vspace{0.2cm}


\noindent
Due to transitions between neutrino flavours (${\nu_e}$, ${\nu_\mu}$, ${\nu_\tau}$), the neutrino masses
 are generated and 
the mass eigenstate of the neutrino system  (${\nu_1}$, ${\nu_2}$, ${\nu_3}$)  becomes a superposition of the  flavour eigenstate.
When considering the simpler two flavour ($\nu_\alpha$, $\nu_\beta$) case, the mass eigenstate ($\nu_1$, $\nu_2$) expresses as

\begin{equation}
 \begin{pmatrix}
   \nu_1 \\ \nu_2
 \end{pmatrix}
=
\begin{pmatrix}
 \cos\theta & -\sin\theta \\
 \sin\theta & \cos\theta
\end{pmatrix}
\begin{pmatrix}
   \nu_\alpha \\ \nu_\beta
 \end{pmatrix}.
 \label{eq:Intro:MixingMatrix}
\end{equation}

\noindent
If we started with $\nu_\alpha$, we might observe a $\nu_\beta$ at a certain distance $L$, due to neutrino oscillations.
The probability of the $\nu_\beta$ appearance as a function of the distance 
is 


\begin{equation}
P\left(\nu_\alpha \to \nu_\beta \right) = \sin^2 2\theta \sin^2 \frac{\Delta m^2 L}{4E}, 
\label{eq:Intro:App_Probability}
\end{equation}

\noindent
where $E$ is neutrino energy and $\Delta m^2 = m_2^2 - m_1^2$ is the difference of square of $\nu_2$ and $\nu_1$ masses.
The probability oscillates due to the interference between the amplitudes of propagation;  $(\nu_\alpha \to \nu_1 \to \nu_\beta)$ 
and 
$(\nu_\alpha \to \nu_2 \to \nu_\beta)$. 
The disappearance probability of $\nu_\alpha$ is therefore expressed as
\begin{equation}
P\left(\nu_\alpha \to \nu_\alpha \right) = 1 - P\left(\nu_\alpha \to \nu_\beta \right).
\label{eq:Intro:DisApp_Probability}
\end{equation}

\noindent
The first experimental evidence for neutrino oscillations and, still, most of the information today relies on high precision disappearance measurements with about 50\,years of history.

The establishment of the neutrino oscillation phenomenon~\cite{Ref_SK98, Ref_SNO02, Ref_KL05} came as a solution of  atmospheric and solar neutrino anomalies  around the year 2000.    
Those  results had indicated two consequences: a new oscillation mode, labelled $\mathbf{\theta_{13}}$, and the possibility to observe CP-violation, if $\mathbf{\theta_{13}}$ was sizeable.
The reactor neutrino experiments CHOOZ~\cite{Ref_CHOOZ} and Palo Verde~\cite{Ref_PV} set upper limit of $\sin^22\theta_{13}< 0.15$ already before 2001. 
The Double Chooz (DC) group was formed in 2006 to measure the $\mathbf{\theta_{13}}$ more precisely making use of near and far detector configuration~\cite{DC_prop}.
The DC  experiment has played a pioneering role in this oscillation channel by providing the first positive evidence, in 2011~\cite{Ref_DC}, in combination with the $\nu_\mu \to \nu_e$ appearance results of  T2K~\cite{Ref_T2K11} and MINOS~\cite{Ref_MINOS11} experiments.
The establishment of $\mathbf{\theta_{13}}$ awaited the Daya Bay experiment's observation in 2012~\cite{Ref_DYB}; confirmed soon after by the RENO experiment~\cite{Ref_RENO}. 
Today's world best value~\cite{Ref_PDG} is driven by the statistical combination of the latest published $\theta_{13}$ results~\cite{Ref_Gd-III,Ref_H-III,Ref_DYBGd-Last,Ref_DYBH-Last,Ref_RENO-Last}.
A reassuring feature for the field is that all reactor \ang\ experiments are redundant
which is critical to ensure a robust and unambiguous result.
Therefore, multi-experimental validation framework is highly beneficial.
A working group formed by all three experiments is on-going with the goal to assess both internal consistency and coherent systematics treatment of each experiment.

Besides reactor experiments, neutrino beam experiments, such as T2K~\cite{Ref_T2K-Last}, NOvA~\cite{Ref_NOvA-Last} and MINOS~\cite{Ref_MINOS-Last}, are also sensitive to \ang\ via the sub-dominant appearance $\nu_{\mu} \to \nu_e$ and $\bar\nu_{\mu} \to \bar\nu_e$ oscillation modes.
However, their ability for a high precision measurement of \ang\ is limited by uncertainties and unknowns such as the $\delta_{\rm CP}$ and  
$\theta_{23}$ octant degeneracy.
Conversely, this channel allows them to explore neutrino oscillation CP-violation directly.
Currently, the beam experiments provide the first CP-violation explorations~\cite{Ref_T2K-Last, Ref_NOvA-Last} by using the value of \ang\ from reactor experiments as input.
The latest data allow beam experiments to obtain the first hints of non-zero $\delta_{\rm CP}$.
This result embodies a remarkable demonstration of synergy and compatibility across both reactor and beam measurements.

In this article, DC reports its multi-detector results for the first time with its 4$^{th}$ data release comprising 865 days exposure.
DC measures reactor $\overline{\nu}_e$ coming from EDF company Chooz twin reactors by identical near and far neutrino detectors.
The \ang\ value was extracted from the disappearance and spectrum distortion of the reactor $\overline{\nu}_e$ caused by the baseline difference of the two detectors.
Most of the systematic uncertainties are cancelled out by using functionally identical detectors observing the same reactor $\overline{\nu}_e$ sources.
DC can provide clean analysis results by using several unprecedented techniques 
such as the exploitation of the effective iso-flux site geometry which cancels possible difference between the two reactors, a model-independent background estimation via reactor power modulation including reactor-off data 
 and the total neutron capture detection technique which significantly increases the neutrino event statistics. 
 Those analysis details are also explained. 
In addition, the DC near detector is used to characterise the rate and shape difference between data and predicted spectra,
and 
measure the most precise neutrino flux to date, given by the mean cross-section per fission $\langle \sigma_f\rangle$.
The latter can be used as reference in other reactor neutrino experiments for an accurate neutrino event rate prediction.
\begin{figure}[t]
	\centering
	\includegraphics[scale=0.2]{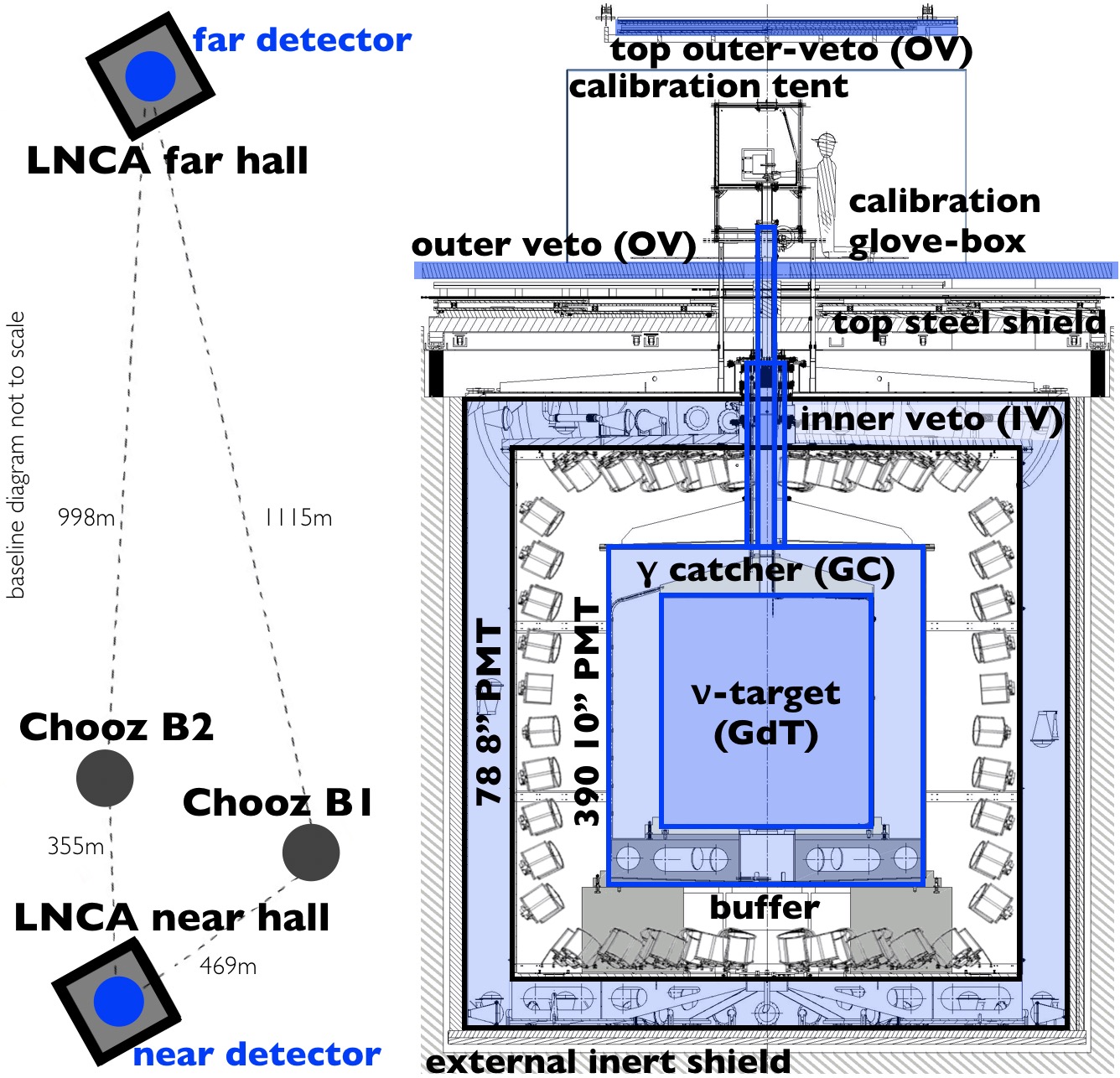}
	\caption{ \small 
	{\bf Laboratory and DC Detectors.}
	The underground LNCA (Laboratoire Neutrino Champagne-Ardenne) site allows for an almost iso-flux geometry ({\bf left}) to the two identical DC detectors ({\bf right}) yielding major inter-detector cancellation of reactor flux and detection systematics.
	Active BG rejection is achieved by the exploitation of the multi-layer (blue shaded) liquid-scintillator design
	whose light is read out by many photomulipliers via a Flash-ADC  deadtime-less electronics
	The {``Inner Detector''} ({ID}) is subdivided into 3 optically coupled volumes: 
	i) $\nu$-Target ({GdT}, 10~m$^3$ liquid scintillator Gd 1~g/l loaded), 
	ii) $\gamma$-Catcher ({ GC}, 23~m$^3$ liquid scintillator), 
	and 
	iii) buffer (100~m$^3$ non-scintillating oil).
	The ``Inner Veto'' ({IV}, 0.5~m thick liquid scintillator) fully surrounds the ID while the ``Outer Veto'' ({OV}, tracking plastic scintillator strip) is placed on the top.
	The IV tags external rock $\gamma$'s (anti-Compton veto), fast-neutrons and cosmic $\mu$'s while the OV only sees cosmic $\mu$'s, covering the ID chimney region.
	An external inert shield surround the IV: 15~cm steel (FD) and 1~m water (ND).
	The glove-box allows clean and safe deployment of the same calibration sources ($^{252}$Cf, $^{60}$Co, $^{68}$Ge, $^{137}$Cs) in both the ND and FD.
	}
	\label{Fig-Det}
\end{figure}

\section*{\label{sec:technique}The Double Chooz Experiment}

\noindent
The DC experiment relies on two identical detectors~\cite{Ref_Det} 
and two of the most powerful pressurised water reactors of the N4 plant series, whose full power is 8.5~GW thermal power (i.e. $\sim$10$^{21}$~$\bar\nu_e$/s flux).
The near (ND) and far (FD) detectors are located at the respective average distance of $\sim$400~m and $\sim$1050~m to the Chooz reactors (B1 and B2).
The \ang\ signature manifests itself as a rate deficit with an up to $\sim$10\% spectral distortion in the FD relative to the almost undistorted ND spectrum.
Thus, DC performs a ``rate+shape'' \ang\ measurement whose statistical uncertainty is dominated by the FD.
There are three types of sources of systematic uncertainties:
{\it detection} (including the estimation of the neutrino energy) and {\it background} are internally constrained with DC data, 
while the {\it reactor flux} relies on an external reactor model.
The model is commonly used by most reactor experiments while here it is customised to the specific DC reactor conditions.
The simple Chooz multi-reactor site geometry enables to place the ND at the effective iso-flux\footnote{The detector locations are slightly off from iso-flux while this has negligible impact.} position relative to the FD.
This implies meeting the condition L$_{\mbox{\tiny B1-ND}}$/L$_{\mbox{\tiny B1-FD}}$ $\approx$ L$_{\mbox{\tiny B2-ND}}$/L$_{\mbox{\tiny B2-FD}}$ for each reactor-detector pair distance (L).
This way, both the FD and the ND are exposed to both reactors
with the same fraction.
From the single-detector (SD) to the multi-detector (MD) configurations, major systematics cancellation occurs by virtue of correlations due to identical detectors (detection systematics) and the iso-flux reactor geometry (flux systematics).
The site and detectors are briefly described in Fig.~\ref{Fig-Det} -- see Appendix for details.
In this release, 481~days of data from single detector operation (FD-I, April 2011 until January 2013) previous to commissioning of the ND and 384~days of data with both detectors FD and ND (FD-II, January 2015 until April 2016) are combined. 
The result presented here supersedes our previous~\cite{Ref_Gd-III,Ref_H-III}.

Each N4 reactor typically runs at maximum power allowing the lowest power uncertainty (0.5\%), or else they stop a few weeks once per year to refuel.
The Chooz total reactor power modulation allows for ``2-reactors'' (both on), ``1-reactor'' (either on) and the unique BG only~\cite{Ref_DCReactorOFF}  ``0-reactor'' (both off) data sets.
An exposure of $\sim$25~days of 0-reactor data is available.
Past FD-I SD results~\cite{Ref_Gd-III,Ref_H-III} employed the Bugey4 experiment data~\cite{Ref_Bugey4} to compensate the ND absence, improving the overall systematics.

\begin{table}
\begin{center}
\begin{tabular}{c|c|cc}
 & Single reactor & \multicolumn{2}{c}{Inter-Reactor} \\ & uncertainty (\%) & SD & MD \\
\hline
Spectrum & $2.19 / 0.06^{\rm B4}$ & Corr. & Corr. \\
Bugey4 $\langle \sigma_{f} \rangle$ & - / $1.41^{\rm B4}$ & Corr. & Corr.\\
Fission Fractions & $0.69 / 0.78^{\rm B4}$ & Corr. & Uncorr. \\
Thermal Power & $0.47$ & Corr. & Uncorr.\\
Energy per Fission & $0.16$ & Corr. & Corr. \\
\hline
Total & $2.27 / 1.68^{\rm B4}$ & &
\end{tabular}
\end{center}
\caption{{\bf Reactor Flux Uncertainties on the Signal Normalisation.} 
Both rate and shape flux uncertainties are treated via covariance matrices as predicted by the data-driven reactor flux model~\cite{Ref_Huber,Ref_Moller,Ref_Hagg} used by DC. The Bugey4 experiment (B4) provides an independent rate constraint via its $\langle \sigma_{f} \rangle$ and therefore extra precision via the cancellation of the common spectrum terms. The uncertainty coming from the reactor-detector baselines is negligible ($<0.01\%$). The unknown inter-reactor correlations are assumed to be correlated for the reactor power $( P_{\rm th} )$ and the fission fractions $( \alpha_{f} )$ in the SD case and uncorrelated for any MD configuration in general (combined uncertainty of $P_{\rm th}$ and $\alpha_{f}$ is $0.83\%$). These assumptions are made to minimise the $\theta_{13}$ sensitivity to be conservative. In the DC case with two reactors the uncertainties on $P_{\rm th}$ and $\alpha_{f}$ are reduced by about a factor of $\sqrt{2}$. Only the uncorrelated terms are relevant for the specific MD case in DC (ND/FD-I and ND/FD-II).}
\label{Tab_ReactorSyst}
\end{table}

The ND monitors the rate+shape of the flux, thus reducing the \ang\ uncertainty from most of the reactor physics and running configuration.
The ND is a direct reactor monitor of the FD-II (iso-flux) and indirect to the FD-I.
The iso-flux implies that the neutrino fluxes are expected to be largely correlated across detectors,
with negligible impact from reactor power or composition variations.
This correlation translates into an almost total rate+shape flux-error cancellation~\cite{DC_prop, Ref_IsoFlux}; a unique DC feature as compared to other reactor-\ang\ experiments~\cite{Ref_PIsoFlux}.
Instead, the FD-I benefits only from partial error cancellation.
The ND provides the reference oscillation spectrum for both FD-I and FD-II for the \ang\ measurement. The shape differences between the FD-I and ND due to the different fuel composition are less than 0.5\%.
The flux systematics reduce from 2.27\% (1.68\% with Bugey4 data -- see Table 1) in the SD case to $\leq 0.83$\% in MD configurations. The uncorrelated uncertainty between the ND and FD-I predictions is estimated to be 0.66\%. It strongly reduces to $\sim 0.1$\%  between the ND and FD-II predictions because of the iso-flux configuration and simultaneous data taking.
In brief, the unique DC geometry grants a framework for a high-precision measurement by cancellation of the flux systematics.

Despite the ND, the reactor $\bar\nu_{e}$ prediction model remains an important element to the analysis, key for the FD-I. 
The same strategy employed in past publications~\cite{Ref_Gd-III} is adopted.
An external reactor $\bar\nu_{e}$ prediction model is used~\cite{Ref_Huber, Ref_Moller} where $^{235}$U, $^{239}$Pu, $^{241}$Pu fissile isotope contributions rely on ILL data~\cite{Ref_ILL}.
DC uses a measurement~\cite{Ref_Hagg} for the $^{238}$U prediction, while Daya Bay and RENO use summation methods.
Dedicated Chooz reactor simulations provide the fission fraction ($\alpha_f$) evolution considering the thermal power data (P$_{\mbox{\tiny th}}$) and re-fuel inventories.
Bugey4 data aid both to constrain the predicted rate~\cite{Ref_2010MCSpF} and to improve the precision.

\begin{figure}[]
	\centering
	\includegraphics[scale=0.47]{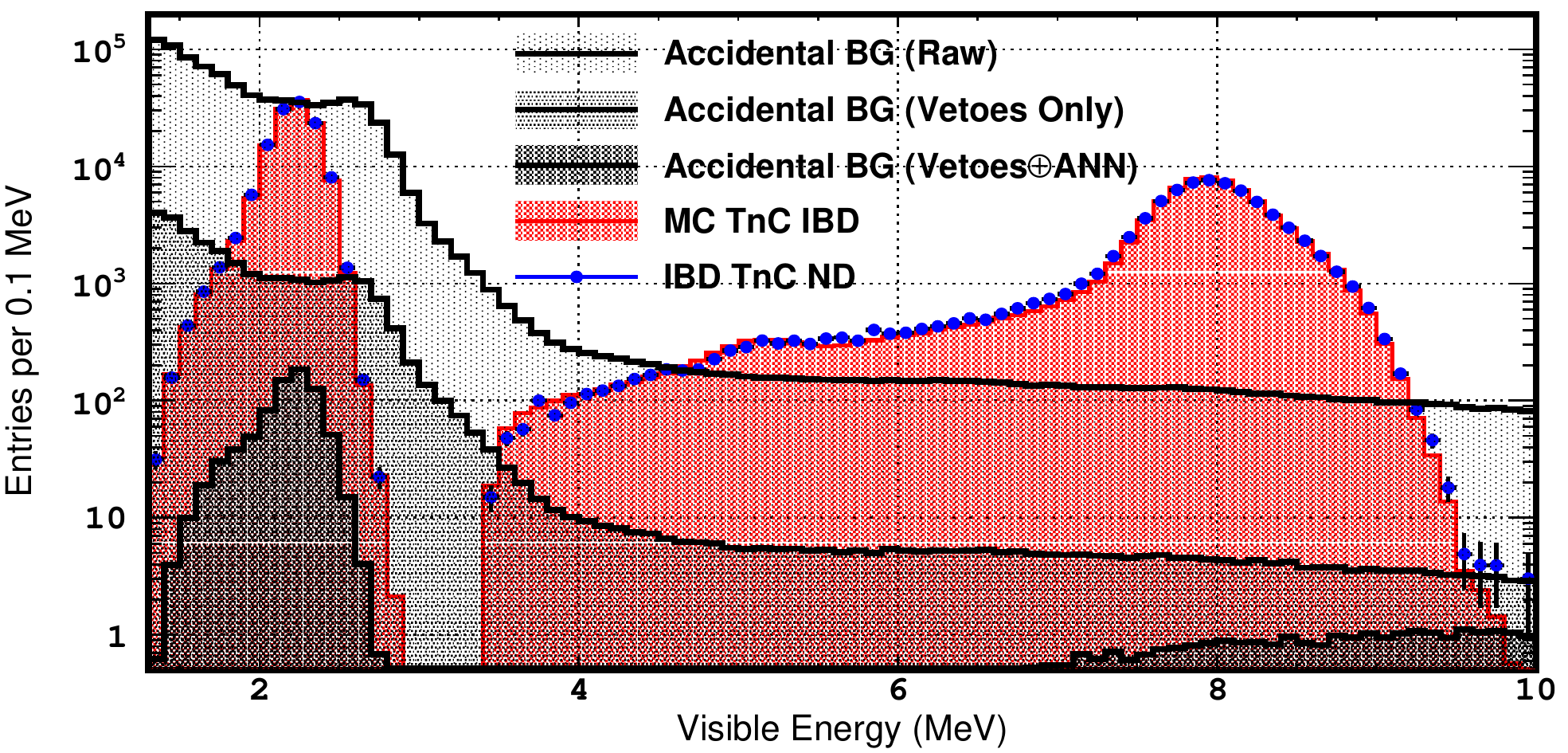}
	\includegraphics[scale=0.48]{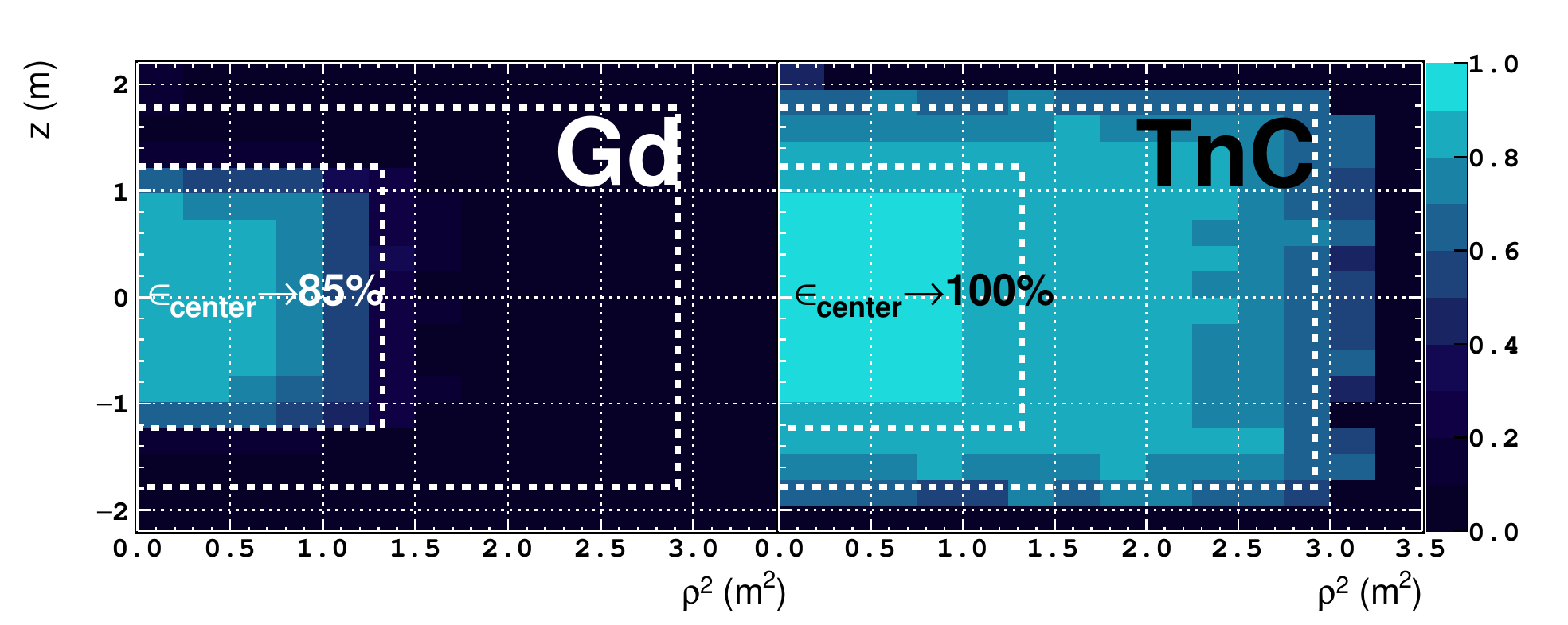}
	\caption{\small 
	{\bf The TnC Detection Principle}.
	The IBD acceptance criteria are widely opened to integrate over all capture-$\gamma$'s: $\sim$2.2~MeV (H-n), $\sim$5.0~MeV (C-n) and $\sim$8~MeV (Gd-n).
	The overwhelming accidental BG (ND) is rejected over \textgreater 4 orders of magnitude below 3.5~MeV ({\bf top}) using vetoes and the ANN selection.
	Excellent data (blue points) to MC (red area) agreement is found in the delayed energy distribution after the rejection.
	The energy scale uncertainty has negligible impact (\textless0.05\%) due to 1.3~MeV cut.
	 The selection efficiency ({\bf bottom}) of the Gd-only~\cite{Ref_Gd-III} ({\bf left}) confines IBD's to the Gd presence.
	 The TnC ({\bf right}) yields detection in the full volume; i.e. GdT (\textgreater 95\%) and GC (\textgreater 80\%).
	 The relative yields are $\sim$61.3\% H-n, $\sim$38.2\% Gd-n and $\sim$0.5\% C-n.
	\label{Fig-TNC}
	}
\end{figure}

\section*{\label{sec:gd++}Neutrino Detection by Total Neutron Capture}

\noindent
Since the discovery of the neutrino, reactor $\bar\nu_e$ are typically detected via the inverse-beta-decay (IBD: $\bar\nu_e + p \to n +  e^+$)~\cite{Ref_IBD} interactions on free protons (i.e. H nuclei) via a coincidence technique, where the \emph{prompt} trigger (e$^+$) is followed by the \emph{delayed} trigger (neutron capture) several tens of $\micro$s later.
The mean capture time $\tau^{\mbox{\tiny capture}}$ is $\sim$200~$\micro$s in a metal-free organic liquid scintillator.
The coincidence aids IBD identification as DC e$^+$ recognition is impractical relative to radiogenic (e$^-$, $\gamma$, $\alpha$) or cosmogenic (p recoil or cosmic $\mu$'s) BGs.
In DC, the Gd is employed via scintillator loading (1~g/l).
Gd's high n-capture probability reduces the mean capture time ($\tau^{\mbox{\tiny capture}} \approx30~\micro$s) and provides a unique n-capture tag ($\sim$8~MeV total energy) allowing for major BG rejection.

Another IBD detection approach opens with the {\it Total Neutron Capture} (TnC) technique presented here for the first time.
The TnC relies on a larger delayed energy range integrating over the $\gamma$-peaks of all capturing elements available, H-n, C-n and Gd-n shown in Fig.~\ref{Fig-TNC}.
Thus, TnC combines past Gd-only~\cite{Ref_Gd-III} and H-only~\cite{Ref_H-III} selections.
The main challenge is the control of larger BGs.
The IBD space-time coincidence definition relies on a multi-variable ANN (Artificial Neural Network) thus rejecting random (uncorrelated) BG coincidences -- see Appendix for details.
More than two orders of magnitude of
accidental background rejection is possible while keeping high the average selection efficiency
at $86.78\pm0.21$\% 
(MC: $86.75\pm0.01$\%) 
and at $85.47\pm0.08$\%
(MC: $85.54\pm0.02$\%) averaged over the prompt energy spectra of the FD and ND, respectively.
The selection efficiency is defined as the inclusive ratio of IBD candidates with the standard and loose ANN cuts.
Thus, the denominator integrates over $\sim$98\% of the detectable IBD's.
The $\sim$1.3\% ND to FD difference, matched by the MC within 0.1\%, is due to the ANN definition which is slightly different to ensure the prompt energy selection efficiency is identical across detectors.

The novel TnC has several remarkable features.
Mainly, the TnC integrates all n-capture elements, thus the $\bar\nu_e$ detection is independent from specific capture details.
This implies that the detection volume expands to 
both the GdT (Gd-target) and GC (gamma-catcher), as shown in Fig.~\ref{Fig-Det},
so the TnC volume increases $\sim$3$\times$ as compared to Gd-only.
This boost in statistics\footnote{$\sim$900~IBD/day (ND) and $\sim$140~IBD/day (FD) with 2 reactors on.} is critical for the DC sensitivity.
In addition, the statistical limitation of the selection systematics is reduced 
as the selection efficiency per volume increases close to 100\% in GdT and $\sim$80\% in GC due to the ANN accidental rejection.
The wider TnC acceptance integrates over most MC inaccuracies, including the complex spill-in/out effect across the GdT-to-GC boundary.
The only relevant boundary is the simpler GC-to-buffer with only H-C on both sides.
So, TnC matches MC better while all element-dependent terms, such as Gd or H fractions, are irrelevant. 
A small concentration of Gd was found inside the GC of the ND due to leakage from the GdT. The described element independence also makes the TnC leak insensitive demonstrating selection stability within 0.1\% in both the ND and FD.

\begin{table}[t]
\centering
\begin{tabular}{ L{3.2cm} | C{2.4cm} | C{2.2cm} }
{\bf Uncertainty (\%)} & {\bf SD} & {\bf MD}\\
\hline
Proton Number      					& 0.65 & 0.39\\
IBD Selection						& 0.33$^{\mbox{\tiny FD}}$ / 0.12$^{\mbox{\tiny ND}}$	& 0.27\\
Boundary Effect 	   					& 0.20 & --\\
Vetoes Efficiency 						& $\leq$0.05 & $\leq$0.05\\
\hline
\end{tabular}
\caption{\small
	{\bf Detection Uncertainties.} 
The central column shows the uncertainties on the signal normalisation for the single detector (SD) case. The multi detector (MD) case in the column on the right shows the uncertainty on the ratio of the signal rates (FD/ND) assuming simultaneous operation (not fully representative for DC fit configuration). 
The total systematics is dominated by the uncertainty of the number of protons for IBD interactions (mainly the GC). This is to be re-measured upon future detector dismantling. The TnC reduces selection systematics as compared to the element dependent detection, since it is not sensitive to the knowledge of the Gd/H fraction of neutron captures. Boundary systematics relying on the modeling of spill-in/out events at volume interfaces in the MC are assumed fully correlated between detectors. The selection systematics rely on an IBD data-driven method, thus inclusively accounting and averaging over selection and energy scale (stability, uniformity and linearity) variations. The vetoes play a negligible role as they were optimized to maximize the selection efficiency while adding a negligible systematic.
\label{tab:detection}
	}
\end{table}

All detection systematics are summarised in Table~\ref{tab:detection} for both SD and the ideal MD case with ND and FD taking data simultaneously.
The \ang\ uncertainty is dominated by the uncertainty on the number of protons in the target/detection volume (or ``proton-number'').
The higher GC proton-number uncertainty (1.1\%) as compared to the GdT ($\sim$0.3\%) is because at the time of the filling of the detectors we did not consider that precise IBD detection in the GC was possible.
The selection systematics estimation uses an IBD data-driven inclusive approach.
Thus, the estimator simultaneously integrates and averages over
a) IBD spectrum and volume,
b) ANN selection correlations and dependences,
c) energy scale systematics including uniformity, stability and linearity and
d) any correlation among all the above terms.
The robustness of the IBD-based methodology was demonstrated with two methods using independent data: 
one was based on $^{252}$Cf data sampling in the GdT~\cite{Ref_Gd-III} while 
the other was based on fast-neutrons data in both the ND and FD over the full volume.
No deviations of more than 1$\sigma$ (respectively 0.1\% and 0.3\%) were observed. 
Lastly, the TnC selection was challenged to per mille precision by estimating the known $^{252}$Cf neutron multiplicity~\cite{Ref_CfMultiplicity}. 
Agreement across ND and FD is within 0.1\% (1$\sigma$).
Thus, the uncertainty on the selection efficiency is demonstrated \textless0.3\% (MD) and \textless0.4\% (SD) excluding the dominant proton-number uncertainty. The proton-number will be re-evaluated with higher precision upon detector dismantling.
In brief, the TnC technique provides a robust IBD detection framework with better selection systematics for both SD and MD physics.

\subsection*{\label{sec:gd++bg}The IBD Signal \& Backgrounds}

\noindent
The BG consists of all physical events mimicking the IBD time-space coincidence implied by the TnC selection.
This includes accidental and correlated BGs.
BG rejection is much harder whenever there is a neutron in the final state.
Due to the small overburden\footnote{ND and FD are, respectively, at $\sim$30~m and $\sim$100~m rock overburden depth, so their cosmic $\mu$ rates are, respectively, $\sim$240~s$^{-1}$ and $\sim$45~s$^{-1}$.}, cosmogenic BGs are dominant: fast-neutrons and unstable isotopes produced from $^{12}$C spallation, such as $^9$Li.
The signature of fast-neutrons is a recoil on H, as prompt, followed by the n-capture, as delayed.
One or more neutrons can participate in a fast-neutrons coincidence.
$^9$Li undergoes a $\beta$-n decay (including $\alpha$'s), but no indications of $^8$He production are found~\cite{Ref_DCLi}.
Since the detector chimney is an effective tagging hole to vertical $\mu$'s, there is a potential background when they stop inside the active detector volume.
Both $\mu$ decay at rest (Michel e$^\pm$) and $\mu$ capture have been carefully studied~\cite{Ref_DCMuonCap} and rejected to a negligible level.
Lastly, the accidental BG is caused by two independent events, mainly from radioactive isotopes inside or around the detectors.

\begin{table}[]
\centering
\begin{tabular}{ L{2.95cm} | C{2.55cm} | C{2.55cm} }
{\bf Rate (day$^{-1}$)} 			& {\bf FD} 			& {\bf ND}\\
\hline 
{\bf IBD Candidates} 						& 112 & 816\\
\hline
{\bf Breakdown}& &\\
 Accidental 				& 
${4.13\pm0.02}$ 	& ${3.110\pm0.004}$ \\
Fast-Neutron				& ${2.50\pm0.05^{\mbox{\tiny}}}$ 	& ${20.85\pm0.31^{\mbox{\tiny}}}$ \\
$^9$Li Isotope			& ${2.62\pm0.27^{\mbox{\tiny}}}$ 	& ${14.52\pm1.48^{\mbox{\tiny}}}$ \\
\multicolumn{1}{ r | }{[$\mu$-tag]}	& $3.01\pm0.60^{\mbox{\tiny}}$ 	& $12.32\pm2.01^{\mbox{\tiny}}$ \\
Stopped-$\mu$ 			& \textless0.19\,@\,98\%CL	& \textless0.21\,@\,98\%CL \\
Others ($^{12}$B, BiPo) 	& \textless0.01 	& $0.04\pm0.01$\\
\hline
{\bf Total}&&\\
$\Sigma$-Exclusive 			& ${9.3\pm0.3^{\mbox{\tiny}}}$ 	& ${38.5\pm1.5^{\mbox{\tiny}}}$ \\
Inclusive (17\,days)			& $9.8\pm0.9$	& $39.6\pm2.5$ \\
\hline
{\bf Signal to BG} &11.0 & 20.2\\
\hline
\end{tabular}
\caption{\small {\bf IBD Candidates Background.}
The rate+shape \ang\ extraction depends on the precise knowledge of each (exclusive) BG rate and shape (i.e. spectra) per detector.
The impact of BG in the FD is larger due to the lower signal rate.
The data are consistent with a BG model with three components, shown in Fig.\ref{Fig-Spectra}, accidental, fast-neutron and $^9$Li.
All other BGs are found or made (via vetoes) negligible.
The BG component accuracy can be demonstrated by providing several independent measurements including 0-reactor data.
This is very valuable for the least precise $^9$Li~\cite{Ref_DCLi} rate ($\sim$10\% uncertainty), where an additional measurement is possible via its time correlation to $\mu$'s (``$\mu$-tag'' indicated).
The lower signal to BG ratio is $\sim$11 as compared to the Gd-only ($\sim$25) due to the larger (\textgreater40$\times$) accidentals.
The total BG precision is 3.2\%$^{\mbox{\tiny FD}}$ and 3.9\%$^{\mbox{\tiny ND}}$.
The total BG ($\Sigma$-exclusive) computation implies a model assumption. 
The comparison and agreement found between $\Sigma$-exclusive (model-dependent) and inclusive (model-independent) measurements provide unique validation of the model itself as well as the BG rates.
The inclusive measurement uses $\sim$17~days of 0-reactor data, not used for \ang\ extraction.}
	\label{tab:bg}
\end{table}

An offline 1.25~ms veto is imposed after each tagged $\mu$, thus rejecting the resulting fast neutrons' captures (\textgreater 5$\times \tau^{\mbox{\tiny capture}}_{\mbox{\tiny }}$) and stopping-$\mu$'s.
The live time loss is 5.4\% (FD) and 25.5\% (ND).
DC has developed a multi-veto method yielding large BG rejection: a rejection factor of 200 is reached for the FD and 35 for the ND as compared to simple time coincidence.
The vetoes are defined identically across ND and FD.
Cosmogenic BG rejection relies on direct $\mu$ and/or neutron tagging using 
ID (inner detector), IV (inner-veto) and OV (outer-veto) detectors', as shown in Fig.~\ref{Fig-Det}.

\begin{figure*}[t!]
	\centering
	\includegraphics[scale=0.47]{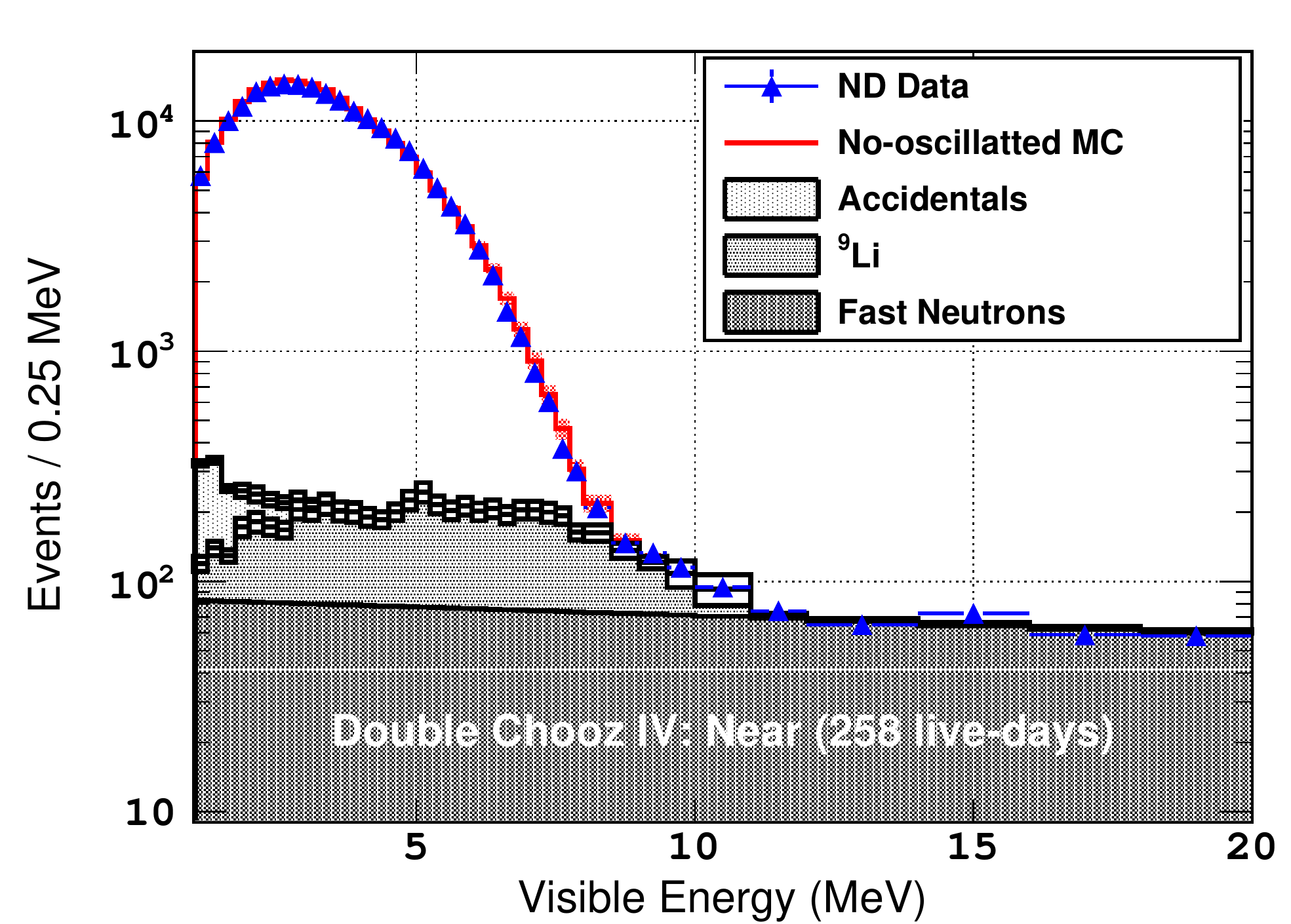}
	\includegraphics[scale=0.47]{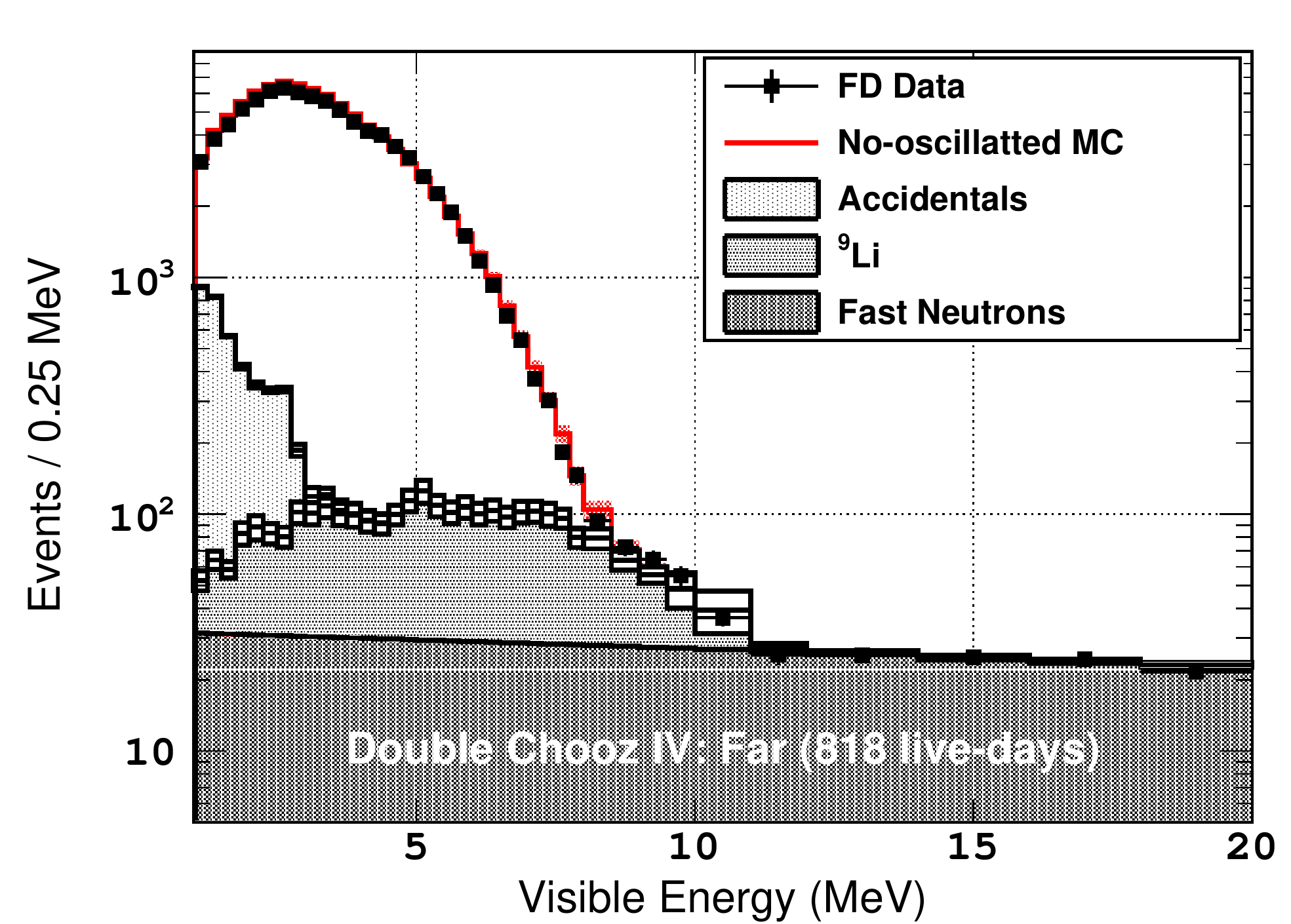}
	\includegraphics[scale=0.47]{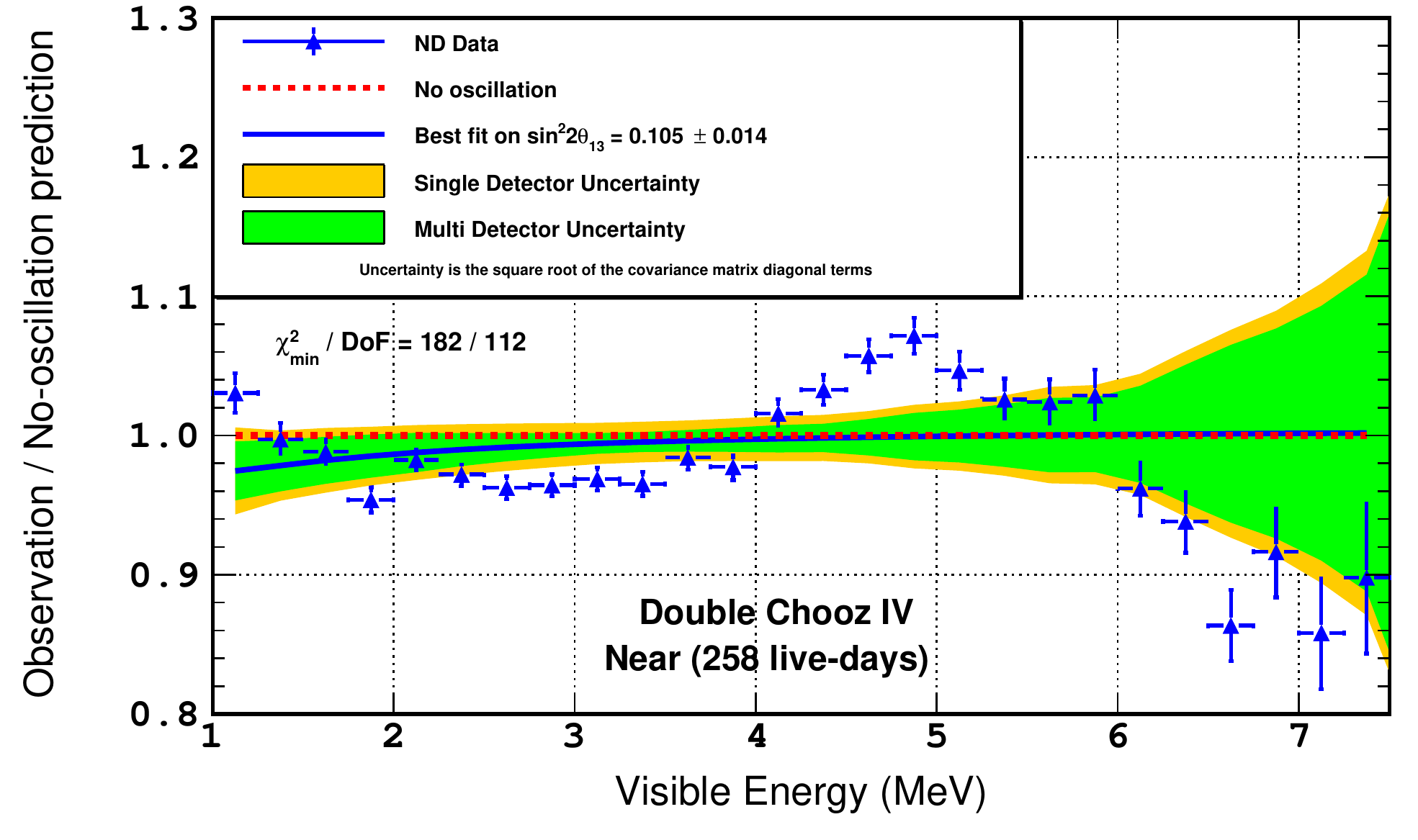}
	\includegraphics[scale=0.47]{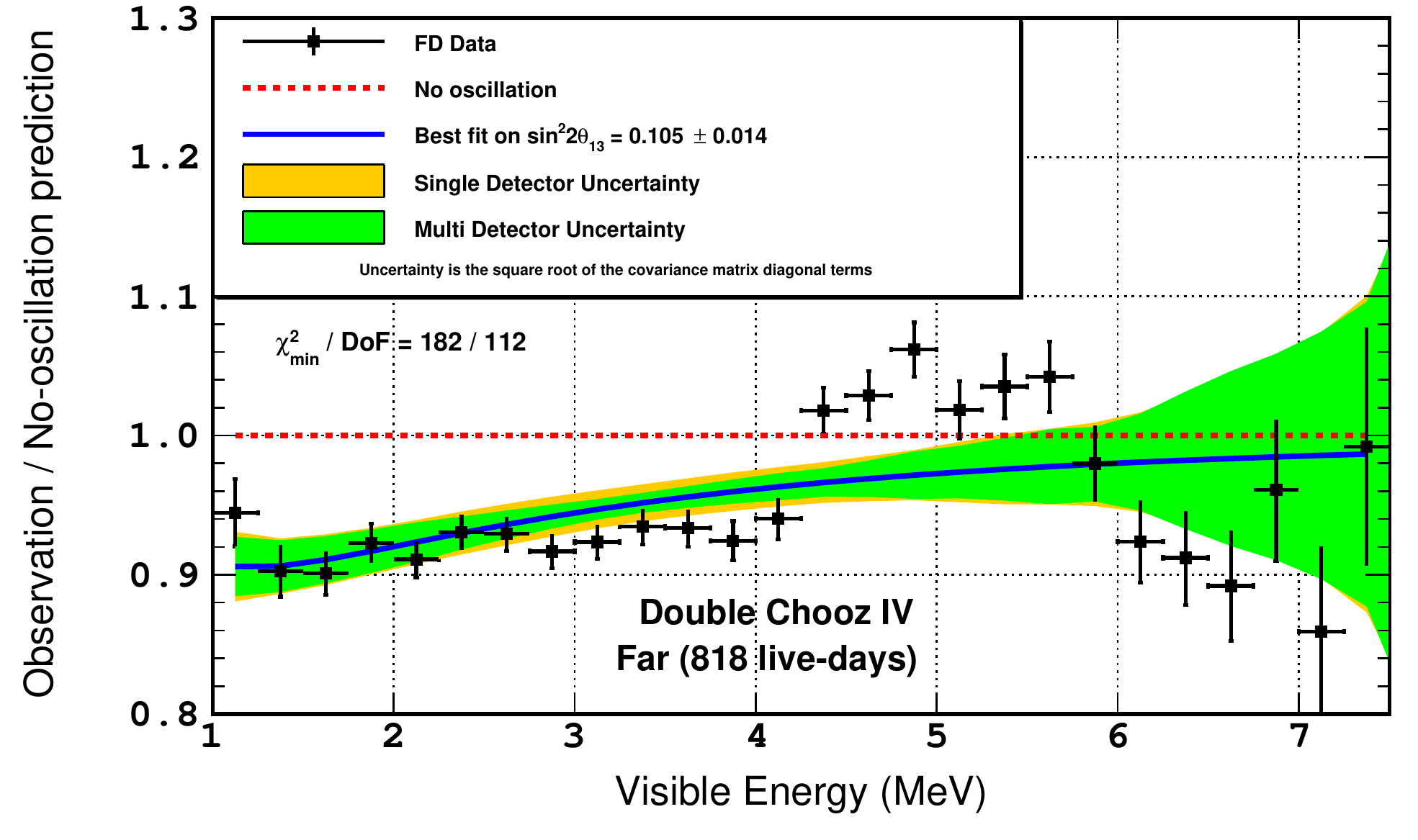}
	\caption{\small 
	{\bf ND and FD Spectra \& SD Ratios}.
	Both ND ($\sim$210k IBD's) and FD ($\sim$90k IBD's) spectra are shown ({\bf top}) within the fit [1.0,20.0]~MeV range, including the un-oscillated MC prediction (red) and the BG model: 
	accidentals (clear grey), 
	$^9$Li (grey) 
	and 
	fast-neutron (dark grey).
	Cosmogenic BGs are estimated during the fit since $^9$Li (unconstraint) dominates in the [7.0,12.0]~MeV region and fast-neutrons above 12~MeV.
	The impact of accidentals to the \ang\ measurement is negligible.
	The data (BG subtracted) to prediction ratio is shown ({\bf bottom}).
	The best fit solution (blue) contrasts with the no-oscillation hypothesis (red).
	Two dominant spectral distortions can be appreciated: 
	the \ang\ signature (mainly FD) 
	and 
	a common 5~MeV excess, leading to a large $\chi^2$/DoF of 182/112.
	Bugey4 constrains the prediction rate. The normalisation with this constraint is lower as compared to the prediction rate not using the Bugey4 information.
	The cancellation of both common distortions and correlated uncertainties takes place from the SD (yellow) to the MD (green) configurations.
	The covariances used (not shown) play an important role during the fit.
	}
	\label{Fig-Spectra}
\end{figure*}

A fraction of $^9$Li and most $^{12}$B are tagged by identifying spallation activity via correlated neutrons upon each tracked $\mu$~\cite{Ref_DCLi}.
Fast-neutron rejection exploits direct neutron tagging in the IV or the primary $\mu$ using the OV.
The ANN discards most accidentals with a rejection factor \textgreater300.
Since IBD's extend to the GC, about 25\% of external $\gamma$'s are tagged in the IV acting as anti-Compton veto.
Stopped $\mu$'s are fully suppressed based on Michel e$^\pm$ discrimination using information of the ID pulse-shapes and goodness of fit information from the position reconstruction.
Any veto using the ID only affects delayed triggers to prevent any prompt spectral distortion affecting \ang.
The IBD inefficiency for all vetoes combined is $4.5$\% (FD) and $5.7$\% (ND) with negligible (\textless0.05\%) systematics.
See Appendix for further complementary selection and veto details.

Table~\ref{tab:bg} summarises the remaining BG estimates including several independent measurements employed to validate the accuracy.
The BG rates, larger compared to Gd-only~\cite{Ref_Gd-III}, have negligible impact on the determination of \ang. 
The BG subtraction impact is small for signal-to-background \textgreater10 while the dominant statistical BG uncertainty is reduced. 
The vetoed BG samples provide copious data-driven spectra, which contain BG information independent of MC simulations.
The shape of cosmogenic BG is found to be identical across ND and FD within statistical uncertainties.
The non-flat energy spectrum of the fast neutrons was carefully evaluated as it could mimic the \ang\ signature.
Its spectra was characterised over an extended window up to 100~MeV. 
The overall impact of BG on \ang\ is marginal.
The dominant BG systematic is the $^9$Li uncertainty.
The BG model accuracy was scrutinised independently with $\sim$17~days of inclusive 0-reactor data samples in both ND and FD-II.
Thus, these data are not used in the \ang\ fit.
No non-statistical bias or tension (\textless1$\sigma$) is found on the measured BG-model, rates and/or spectral shapes.

\begin{figure*}[t!]
	\centering
	\includegraphics[scale=0.47]{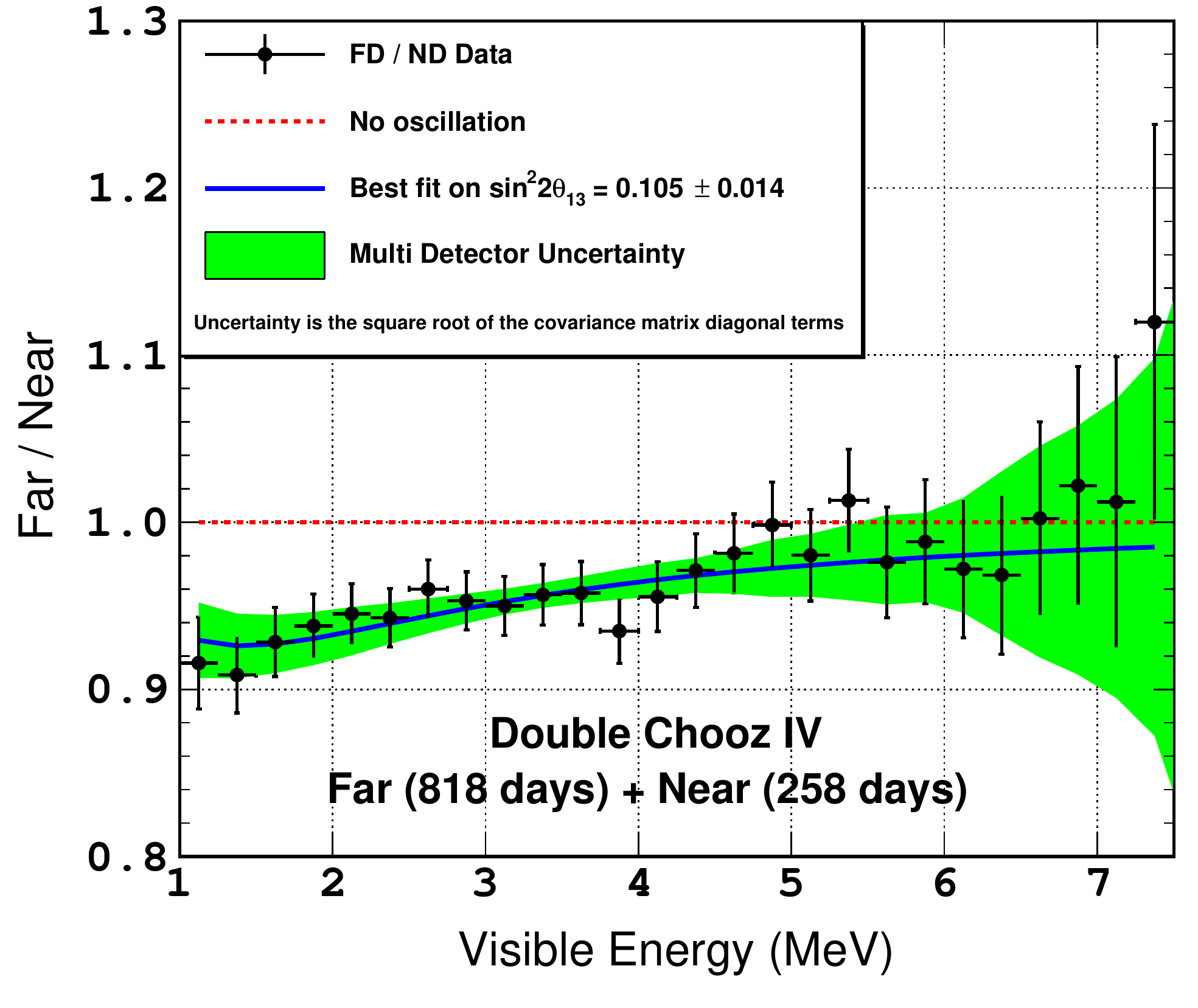} 
	\includegraphics[scale=0.47]{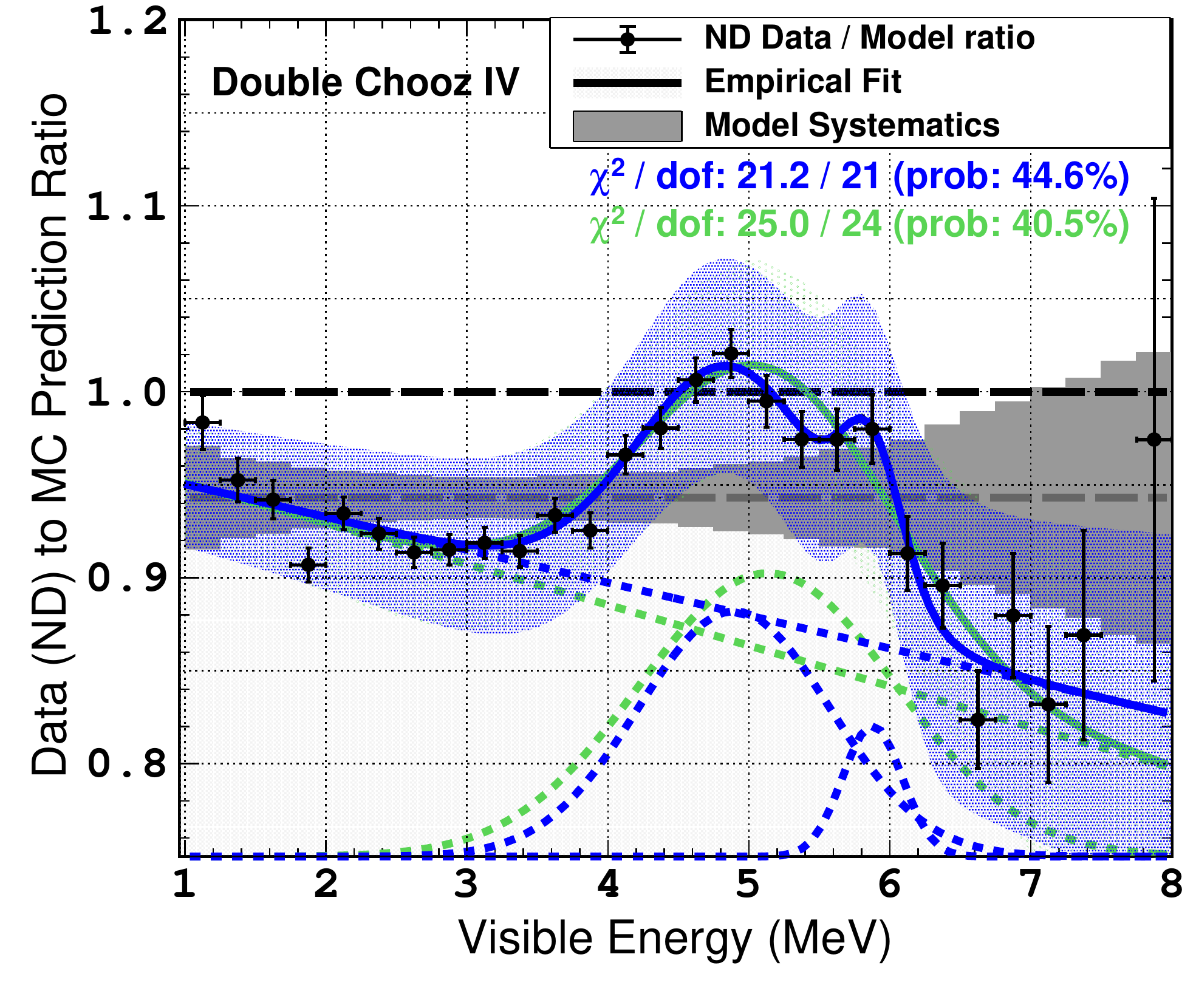}
	\caption{\small 
	{\bf The Spectral Ratios.}
	The FD to ND ratio ({\bf left}) represents a clean $\theta_{13}$ rate+shape disappearance evidence used by the fit for parameter extraction. No traces of any remaining distortion are found, demonstrating the expected inter-detector cancellation key to ensure the $\theta_{13}$ accuracy. Instead, ND data to MC prediction ratio ({\bf right}) allows for precise extraction of the spectral distortion, which is common in FD and ND as seen in Fig.3. Both the rate and shape effects are visible so rate+shape feature extraction is possible. An empirical structure is examined by fitting with two models: one and two empirical Gaussian peaks with a common slope. Both models reproduced data. The origin of those empirical features remains unknown. Shape-only analysis is described in Appendix.
	}
	\label{Fig-Ratio}
\end{figure*}

\section*{\label{sec:t13}The $\mathbf{\theta_{13}}$ Measurement}

\noindent
The \ang\ measurement is obtained by contrasting the observed IBD rate+shape spectral distortion against the specific neutrino oscillation model prediction,
similar to Eq.\ref{eq:Intro:DisApp_Probability},
in which the rate reduces following the flux modulation given by
\begin{equation}
	P(\bar\nu_e \to \bar\nu_e) \approx 1 - \sin^2 2\theta_{13} \sin^2 (1.267 \Delta m^2_{ee} \mbox{L/E}_{\bar \nu_e}) 
\end{equation}
\noindent
where \sang\ is the unknown.
L(m) is the baseline distance between each reactor-detector pair,
E$_{\bar \nu_e}$(MeV) is the neutrino energy obtained from the prompt energy deposition or {\it Visible Energy} (E$_{\bar \nu_e}\approx$ E$_{e^+}$ + 0.78~MeV).
$\Delta m^2_{ee}$ is the pertinent $\nu_e$ weighted average of $\Delta m^2_{31}$ and $\Delta m^2_{32}$~\cite{Ref_DeltaMee}, where $|\Delta m^2_{ee}| = (2.484\pm 0.036)\times10^{-3}$eV$^2$~\cite{Ref_NuFit3.1} is used as input to the fit.

The \ang\ rate+shape fit measurement uses all detectors data simultaneously.
The nominal fit considers the input from each SD fit (data to its MC) including pertinent constraints and correlations.
The SD fit is shown in Fig.~\ref{Fig-Spectra}-(bottom). In our MD analysis, all SD fits (FD-I, FD-II and ND) are simultaneously performed, constrained by the inter-detector correlations such as BG (shape), detection (rate), energy (shape) and flux (rate+shape).
Thus, the common ND provides direct and almost un-oscillated rate+shape reference spectrum. Systematic uncertainties cancel due to correlations with both FD-I and FD-II.
The iso-flux FD-II benefits from the maximum error cancellation.
The \ang\ measurement is, in principle, independent from any common or correlated contributions across the MC and detectors.
Fig.~\ref{Fig-Ratio}-(left) illustrates the inter-detector ratio fit exhibiting the expected \ang\ flux modulation and demonstrating the suppression of the spectral distortion against the MC.
The common MC serves both as link to the neutrino energy (E$_{\bar \nu_e}$) and an inter-detector comparison mediator.
The non-trivial role of the reactor model is scrutinised later on.
This is a delicate point since the data to prediction comparison exhibits clear distortions uncovered by the uncertainties such as the 5~MeV excess shown in Fig~\ref{Fig-Spectra}.
The BG constraints benefit from 
$\sim$8~days of FD-I 0-reactor data (taken in 2011 and 2012) and
the 20~MeV range extension.
The systematics are treated both via covariance matrices (energy and reactor flux) and nuisance pull terms. 
In the covariance treatment data points and uncertainty bands in figures do not fully represent the fit constraints upon minimisation.
Pull terms are also used in the $\chi^2$ minimisation and give access to physical observables (BG rates, inter-detector normalisation, etc) as well as insight to the fit consistency.
There is negligible (\textless1$\sigma$) tension in all fit output values.
The fit strategy follows an unbiasing scheme where performance and robustness to \ang\ measurements are scrutinised using MC and fixed prior to the final data fit.

The best fit value is \sang=$0.105\pm0.014$ (13.3\% precision) with $\chi^2$/DoF~=~182/112 with a $p$-value of 3.2$\times10^{-5}$.
The expected total uncertainty (i.e. sensitivity) was 0.014.
The statistical precision is 0.005, so systematics largely dominate.
The large $\chi^2$/DoF is caused by the mismatch between data and prediction, which is not covered by the model uncertainties. This topic is addressed later on.
The $^9$Li rate is unconstrained in the fit.
This is because the fit $^9$Li sample is up to 50\% statistically correlated to the one used for $^9$Li estimation ($\mu$-to-IBD time correlation), as summarised in Table~\ref{tab:bg}.
The rate+shape spectral distortion is shown in Fig.~\ref{Fig-Ratio}~(right). 
An empirical fit to the spectral distortion residual appears to resolve a structure consistent with a slope and one or two Gaussian peaks. Their origin remains unknown. 
The common normalisation across all detectors can be measured as output of the fit.
The value is $1.004\pm0.008$, while the input value was constraint by the uncertainty of the Bugey4 measurement (1.4\%)~\cite{Ref_Bugey4}. 
The sizeable smaller output uncertainty indicates that DC holds valuable independent information about rate normalisation.

\begin{table}[t!]
\centering
\begin{tabular}{ L{2.2cm} | C{2.8cm} | C{2.8cm} }
{\bf Uncertainty} 		& {\bf Single Syst.} 	& {\bf Total - x}\\
\hline
Reactor Flux 		& 0.0081 (7.6\%)		& 0.0112\\
Detection 			& 0.0073 (6.8\%)		& 0.0113\\
Energy 			    & 0.0018 (1.7\%)		& 0.0121\\
Background 			& 0.0018 (1.7\%)		& 0.0134\\
$|\Delta m^2_{ee}|$	& 0.0018 (1.7\%)		& 0.0140\\
\hline
Statistics 			& 0.0054 (5.0\%)		& --\\
\hline
{\bf Total} & {\bf 0.0141 (13.3\%)} & \\
\end{tabular}
\caption{
	\small 
	{\bf sin$\mathbf{^2{2\theta_{13}}}$ Measurement Uncertainties Breakdown.}
The match between the \ang\ uncertainty from data (0.0139) and the predicted sensitivity (0.0141) allows for a MC uncertainty breakdown. In the central column the fractional uncertainties of the different systematics (x) are given. They were calculated from the sensitivity assuming just one systematic contribution in addition to the statistical uncertainty. The statistical part was then subtracted in quadrature. The total is larger than the square root of the sum of the individual squared uncertainties because of correlations. The difference corresponds to a $(0.0065)^2$ term. The column on the right shows the total uncertainty when the corresponding single systematics is removed. The impact of background and in particular of the energy scale on the sensitivity is higher than one might expect from the values given in the central column. Again, this is due to the correlations.     
	\label{tab:t13sys}
	}
\end{table}

Table~\ref{tab:t13sys} summarises the contributions to the total uncertainty in the oscillation fit.
The FD drives the overall statistical precision ($\sim$90k IBD's).
Because the FD-I data set still represents a large fraction of the data analysed here, the flux systematics are the largest contribution to the uncertainty on \ang. However, the operation of the two detectors in iso-flux configuration during FD-II lead to a large reduction of the systematics and this phase drives the overall sensitivity.
The detection uncertainties are dominated by the large GC proton-number uncertainty.
The BG has a small role, thanks to the statistics and since the $^9$Li rate is constrained in the oscillation fit due to the spectral shape above $\sim7$~MeV.
The non-linearity uncertainty is lower than 0.6\% thanks to the reliable Flash-ADC linearity control.
The impact of deviations from response stability and uniformity is negligible. 
Using the common $^{252}$Cf source fission prompt spectrum, the response linearity was found identical between detectors with no slope greater than 0.1\%.
The impact of the uncertainty on $|\Delta m^2_{ee}|$ is marginal.
Last, some degree of degeneracy among systematics exists and leads to correlations. 
This increases the impact of some terms (see Table~\ref{tab:t13sys}).
To conclude, the reactor (FD-I) and the detection systematics dominate. 
More data are expected to reduce the impact of most systematic errors and correlations between them because the relative impact of the FD-I phase is reduced.
A \textless0.010 precision on the $\sin^2{2\theta_{13}}$ measurement is possible using the full data exposure, if the proton-number systematics were to be improved.

\begin{figure}[t]
	\centering
	\includegraphics[scale=0.47]{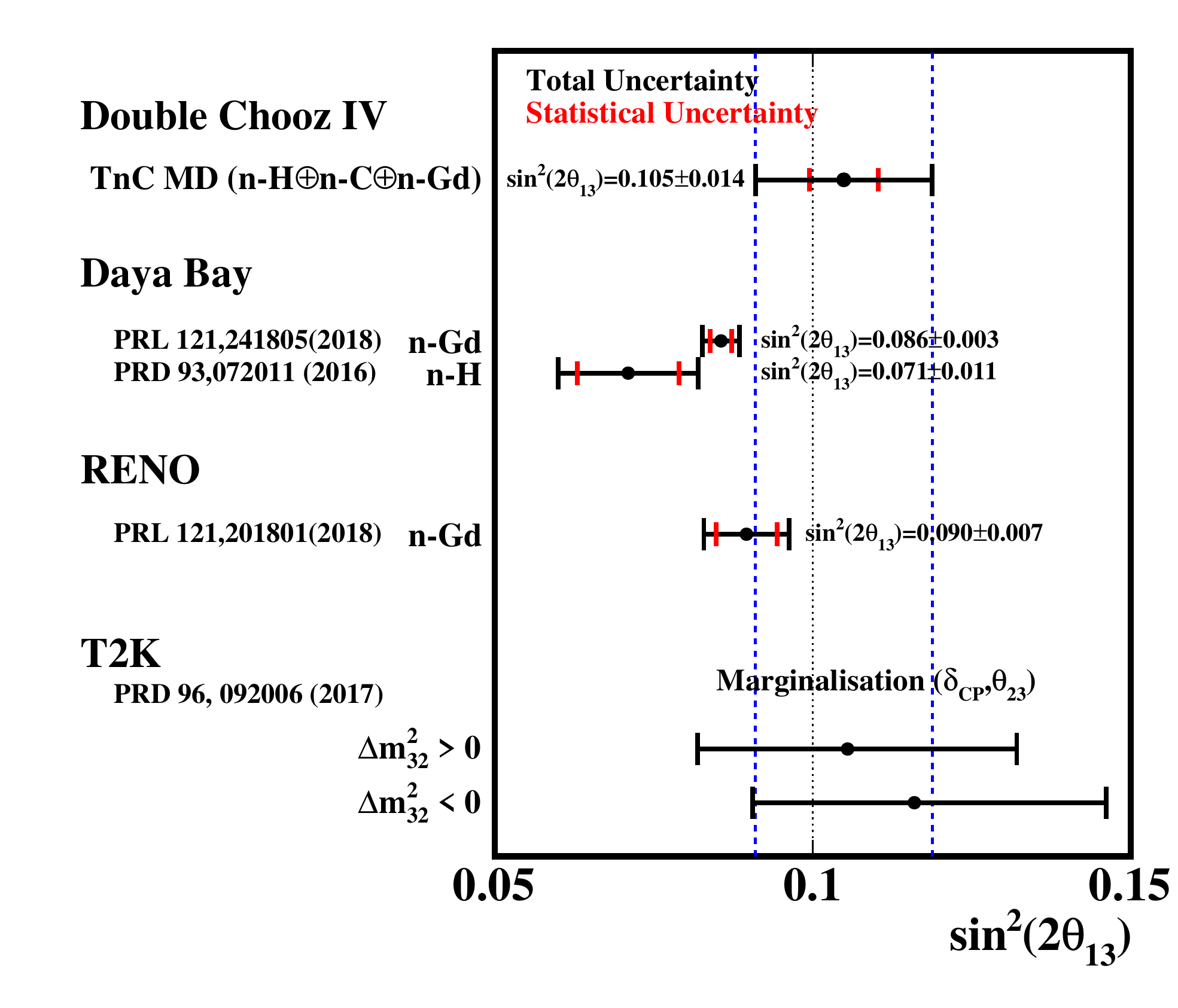}
	\caption{\small 
	{\bf Latest Published $\mathbf{\theta_{13}}$ Measurements.} 
	The most precise published reactor measurements from 
	DC MD TnC (this work),
	DYB~\cite{Ref_DYBGd-Last,Ref_DYBH-Last},
	and RENO~\cite{Ref_RENO-Last} are shown.	
	The latest DC result is consistent with previous ones. 
	Our result exhibits an up  to 48\% higher central value whose significance ranges \textless2.0\,$\sigma$'s.
	The latest T2K~\cite{Ref_T2K-Last} is shown for comparison, whose larger uncertainty considers the marginalisation over the $\theta_{23}$-octant and CP violation.
	}
	\label{Fig-T13W}
\end{figure}

\section*{\label{sec:discussion}Discussion \& Implications}

\noindent
Our reported MD \ang\ exhibits an up to 48\% higher central value whose significance is \textless2.0\,$\sigma$'s compared to all other measurements.
The latest published values of \ang\ are shown in Fig.~\ref{Fig-T13W}.
NOvA~\cite{Ref_NOvA-Last} and MINOS~\cite{Ref_MINOS-Last} are also sensitive to \ang.
Since the statistical uncertainties in reactor experiments are small, a simple statistical fluctuation is unlikely to be the sole cause of the difference.
Differences are however today consistent within the context of the dominant systematics uncertainties.
The consistency among the DC, DYB and RENO reactor measurements remains critical check for the \ang~final global value used everywhere else.

\subsection*{\label{sec:accuracy}Systematic Uncertainty Scrutiny}

\noindent
The reported \ang\ result deserves thorough scrutiny to ensure that the accuracy (i.e. any bias) is controlled well within the quoted uncertainties.
In order to do this, DC has
performed several independent checks.
This is only possible for internal systematics; i.e. those relying on the experiment's data.
This was reported in previous sections for the case of detection, energy and BG systematics.
The case of the reactor flux model is exceptional, as it cannot be tested directly with DC data.
Today's IBD data exhibit a significant discrepancy in terms of both rate (i.e. deficit) and shape (i.e. possible slope and excess around $\sim$5~MeV), as illustrated in Fig.~\ref{Fig-Ratio} -- see Appendix for details.
While there is so far an unsettled debate on its origin~\cite{Ref_5MeVpaper}, here we shall focus on the empirical impact on the \ang\ determination.
Thus, the remaining discussion addresses the subtle role of the reactor model and its systematics on the reported \ang\ measurement.

\begin{figure}[t]
	\centering
	\includegraphics[scale=0.47]{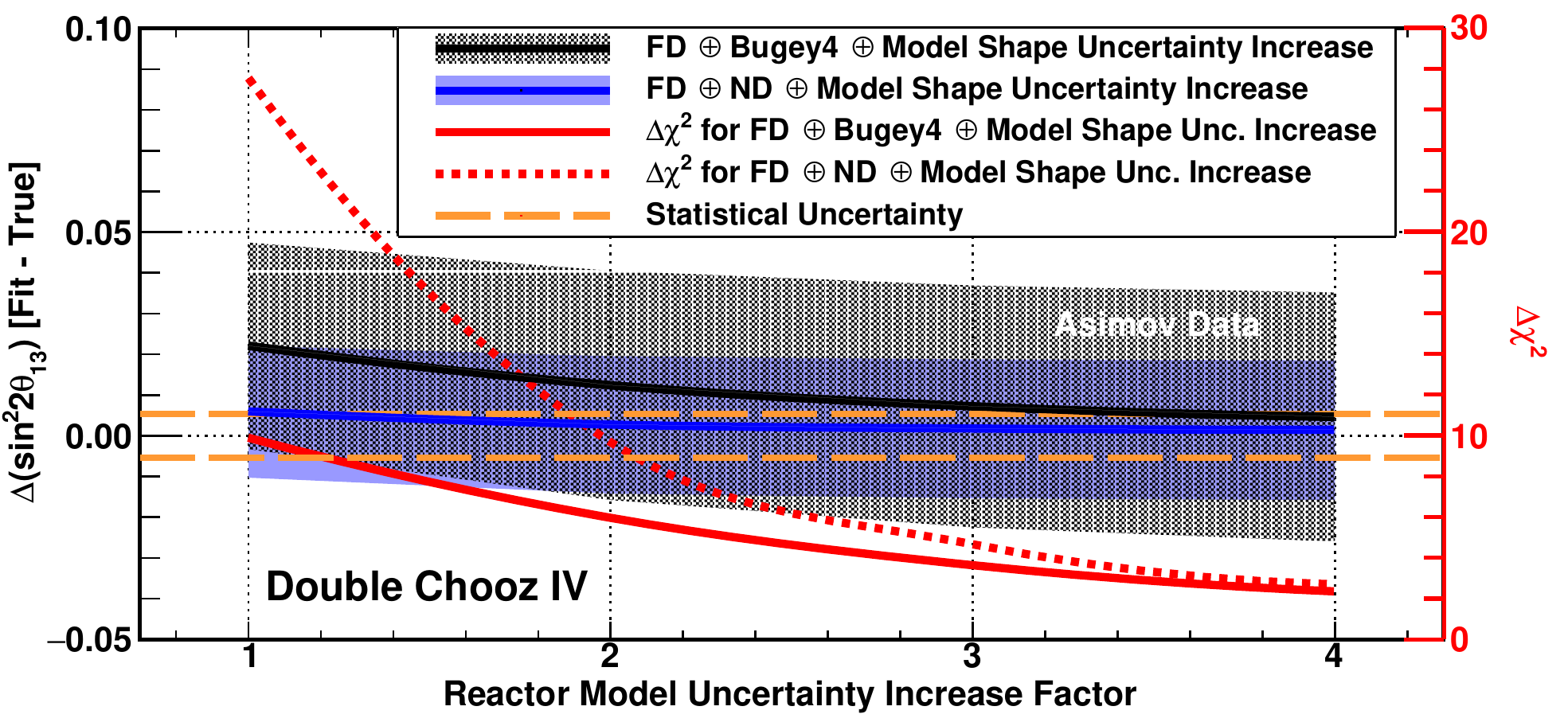}
	\caption{\small 
	{\bf Reactor Model Uncertainty Impact on $\mathbf{\theta_{13}}$.}
	Asimov data is used here to illustrate the spectral distortion impact on \ang\ when considering a similar distortion as that found in data.
	Bugey4 is needed to correct the SD ({\bf black}) rate normalisation. 
	Else, an unbiased \ang\ measurement is impossible, even if arbitrarily increasing the uncertainties.
	The SD \ang\ value exhibits a strong dependence on the shape uncertainty of the reactor model due to the spectral distortion.
	Both SD and MD ({\bf blue}) match to the input \ang\ only when considering an increase of up to 4$\sigma$ of this uncertainty.
	This increase provides an IBD-based empirical new prescription which yields a robust \ang\ measurement in both bases.
	The extra uncertainty accommodates the otherwise unaccounted 5~MeV spectral excess.
	This is corroborated with data (see text). While this behaviour is specific to the \ang\ signature, the main pattern remains: 
	any shape-dependent SD result might be spurious due to the reactor model dependences.
	 The \ang\ from MD data is found to vary \textless 1.0\% 
	 thus demonstrating its better stability.
	The model uncertainty underestimation arises in the MD case mainly via the reported $\chi^2$/DoF tension due to the ND statistical precision.
	This tension vanishes when similarly increasing the model uncertainty.
	}
	\label{Fig-ReactorModel2T13}
\end{figure}

\paragraph{\bf Impact of Reactor Model on \ang.}
Since DC data has a limited ability to test the validity of the model, the stability of the \ang\ measurement is scrutinised and demonstrated against the behaviour of the reactor model for both the SD and MD configurations as explained in Fig.~\ref{Fig-ReactorModel2T13}.

The {\bf SD case} is more illustrative as a stronger dependence on the model biases is expected due to the lack of a ND.
Also, past DC results with data from the FD only can be directly validated\footnote{The risk to compromise SD accuracy was already suspected in~\cite{Ref_H-III}, so a rate-only \ang\ measurement was conservatively adopted as baseline.}.
Indeed, we demonstrate that a measurement of \ang\ is compromised using the standard rate+shape model prescription due to the large data to model mismatch.
Despite the constraint on the rate from Bugey4, the shape distortion biases the fit via the shape-only term.
This effect grows with statistics.
The new empirical prescription for FD-I+FD-II data explained in Fig.~\ref{Fig-ReactorModel2T13} yields 
\sang=$0.108\pm0.028$ ($\chi^2$/DoF = 53/74) 
matching the MD result.
This is the best SD \ang\ measurement to date.
Conversely, the standard reactor uncertainty leads to 
\sang=$0.122\pm0.022$ ($\chi^2$/DoF: 105/74).
Thus, SD is proved a fragile measurement framework due to the unavoidable dependence on the rate+shape reactor model deviations and systematics.

The {\bf MD case} is demonstrated a robust \ang\ measurement.
With the ND, the inter-detector cancellation protects \ang\ largely from any common or correlated rate+shape bias.
The model uncertainty underestimation mainly manifests as the larger $\chi^2$/DoF tension due to the large ND statistical precision.
The increase on the uncertainty of the reference spectrum causes both a more robust \ang\ value (\textless1\% effect)
and the alleviation of the $\chi^2$ tension.
The latter was corroborated with data, in which $\chi^2$ went from 182 to 93 (DoF: 112).
Instead, the increase of the model uncertainty has almost no impact on the \ang\ precision, as shown in Fig.~\ref{Fig-ReactorModel2T13}.

Our studies allow a few empirical observations linked to today's model limitations whose origins remain unknown.
i) The 1$\sigma$ envelope for today's prediction appears insufficient to accommodate the mismatch between data and model for both rate and shape.
A better understanding of the origin of model deviations remains critical.
In the meantime, the adoption of IBD data driven methods is the only way to bypass the model limitations. 
Here DC demonstrates that Bugey4 (or alike) can be used to bypass the rate model bias with few per mille accuracy.
However, the same is less evident for the spectral shape bias or distortion due to unresolved remaining differences among experiments at the few \% level today -- see Appendix. 
ii) the DC IBD data prescription favours the increase of today's shape-only uncertainty to extend the empirical model, as described in Fig.~\ref{Fig-ReactorModel2T13}. 
Unless new physics proves otherwise, significant reactor model progress is needed to attain SD precision below $\sim$6\%.
iii) the spectral-based bias is expected to depend on the relative position between the dominant features, such as the $\sim$5~MeV excess, and the energy range where the \ang~oscillations affect the spectrum.

\noindent
\paragraph{\bf Rate and Shape Decomposition.}
Since the \ang\ measurement exploits both the rate and the shape, further insight comes from splitting the measurement into the rate-only (16\% precision) and shape-only (43\% precision) contributions.
The main \ang\ constraint is due to rate-only (systematics dominated).
The shape-only information has enhanced significantly due to the higher statistics in the FD as compared to previous DC results~\cite{Ref_Gd-III}.
No tension is found between rate and shape measurements (\textless0.5$\sigma$).
The shape-only $\theta_{13}$ fit value is about 20\% lower as compared to our main result whereas the rate-only fit is higher. This indicates shape effects as the observed spectral 
distortions do not introduce a bias towards a higher \ang.
The shape stability of the central value is seen by freeing the $|\Delta m^2_{ee}|$ marginalisation:
\sang=$0.104_{-0.019}^{+0.032}$ and $|\Delta m^2_{ee}| = (2.49^{+0.40}_{-0.49})\times10^{-3}$eV$^2$ are obtained. 
A loss in precision is expected due to the DC non-optimal baseline. 
See Appendix for further cross-checks.

\noindent
\paragraph{\bf FD-I and FD-II Decomposition.}
Thanks to the direct iso-flux monitoring, the reactor flux prediction of FD-II is largely correlated with the ND ($\sim 0.1$\% uncorrelated normalisation uncertainty), hence the precision on \ang\ is much better as compared to \mbox{FD-I}.
It is interesting to decompose the \ang\ measurement into the statistically independent FD-I (22\% precision) and \mbox{FD-II} (15\% precision) samples.
Indeed, FD-II drives the reported central value of \ang.
Using the SD, the FD-I and FD-II samples are demonstrated to be statistically consistent ($\sim$0.7$\sigma$).
FD-II (46\% of data) is expected to dominate future DC results.

\begin{figure}[t]
	\centering
	\includegraphics[scale=0.47]{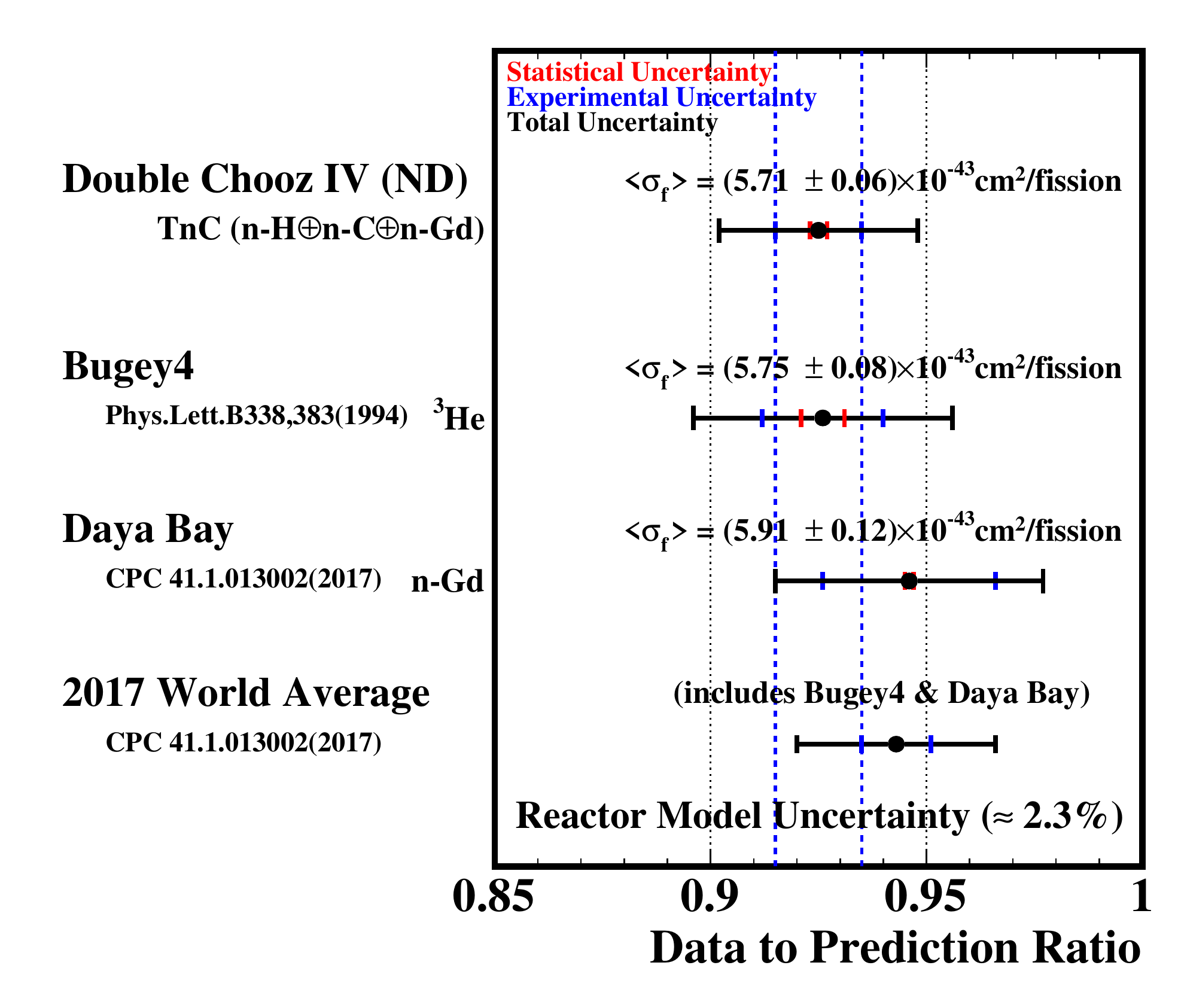}
	\caption{\small 
	{\bf Latest Published $\mathbf{\langle\sigma_f\rangle}$ and R$^{\langle\sigma_f\rangle}$ Measurements.} 
	The R$^{\langle\sigma_f\rangle}$ ratio of ${\langle\sigma_f\rangle}$ (mean cross-section per fission) for   
	DC(ND), 
	Bugey4~\cite{Ref_Bugey4,Ref_DYB-MCSpF}, 
	DYB~\cite{Ref_DYB-MCSpF} 
	and
	the 2017 world average~\cite{Ref_DYB-MCSpF} are shown. 
	The DC shown results are 
	${\langle\sigma_f\rangle}$=$5.71\pm0.06\times10^{43}$cm$^2$/fission
	with a
	R$^{\langle\sigma_f\rangle}$(ND)=$0.925\pm0.002(stat.)\pm0.010(exp)\pm0.023(model)$ upon corrections, 
	including the \ang\ dependence for DC(ND) and DYB.
	The corresponding ${\langle\sigma_f\rangle}$ are also quoted for DC, Bugey4 and DYB.	
	DC ND supersedes Bugey4 as the most precise measurement to date, thanks to the good systematics control using the TnC selection.
	Rate normalisation agreement is found across all experiments, as indicated by a consistent R within uncertainties 
	(black), dominated by the prediction error.
	}
	\label{Fig-MCSpF}
\end{figure}

\noindent
\paragraph{\bf The ND Rate Dependence: $\mathbf{\langle \sigma_f\rangle}$.}
Past DC SD \ang\ results have relied on Bugey4 normalisation constraint.
Beyond the better precision achieved, this implies that those measurements depended on the Bugey4 rate normalisation.
In the MD case, the impact of Bugey4 cancels across both the ND and FD.
The consistency of the ND normalisation can be experimentally tested via the measurement of $\langle \sigma_f\rangle$; i.e. the mean cross-section per fission.
The value of $\langle \sigma_f\rangle$ is proportional to the reactor mean flux.
So, the reactor luminosity can be used as an inter-experiment reference allowing comparison across experiments.
This way, the ND normalisation can be validated against Bugey4 -- the most precise measurement to date and reference to all past DC measurements.
Fig.~\ref{Fig-MCSpF} shows the data to MC model ratio (R$^{\langle\sigma_f\rangle}$)
for the most relevant published results so far and their corresponding $\langle \sigma_f\rangle$.
The R$^{\langle\sigma_f\rangle}$(ND)
is in agreement within 1$\sigma$, dominated by prediction uncertainty, with all other experiments, including the  2017 world average.
The prediction normalisation used~\cite{Ref_DYB-MCSpF} allows easier comparability across experiments -- see further details in the Appendix.
Other normalisation estimations~\cite{Ref_Giunti} are consistent within order 1\%.
The ND measurement of $\langle \sigma_f\rangle = (5.71\pm0.06)\times10^{-43}$cm$^{2}$/fission is the most precise measurement to date.
The $\langle \sigma_f\rangle$ central values of Bugey4 and DC(ND) agree \textless0.5\%, once corrected by the relative differences in average fuel composition.
This was quantified by the overall DC \ang\ fit normalisation output when using the Bugey4 constraint.

\section*{\label{sec:conc}Conclusions}

\noindent
The first DC \ang\ multi-detector (MD) measurement is presented with a best value \sang=$0.105\pm0.014$.
This measurement pioneers the IBD Total Neutron Capture (TnC) detection technique with a major reduction of both SD (single-detector) and MD selection systematics and by a boost of statistics by a factor $\sim$2.5$\times$ relative to Gd-n selection.
DC demonstrates the robustness of the quoted background, detection and energy systematics via the articulation of several independent measurements for each systematic.
The external reactor flux model systematics are studied and shown to have negligible impact for the extraction of \ang\ with the MD configuration.
However, the SD \ang\ extraction is more sensitive
to the systematics of the predicted spectrum. 
An empirical model extension is introduced here with increased uncertainties for the reactor systematics. In this way an accurate SD \ang\ measurement in very good agreement with the MD analysis is demonstrated.
DC also reports here for the first time an empirical model of the distortion found between the measured and the predicted spectra. The observed structure of the empirical fit might shed light on the origin behind these deviations. 
DC also reports here the most precise mean cross-section per
fission $\langle \sigma_f\rangle = (5.71\pm0.06)\times10^{-43}$cm$^{2}$ to date, in
good agreement with Bugey4 and others experiments.
All the results presented here are expected to provide access to the best reactor-\ang\ knowledge and related physics.
The DC \ang\ precision is expected to improve with more statistics and better systematics such as the proton-number.

%


\section*{Acknowledgments}
{\small
This publication is dedicated to our colleague Herv\'{e} de Kerret.
We thank the EDF (``Electricity of France'') company; 
the European fund FEDER; 
the R\'egion Grand Est (formerly known as the R\'egion Champagne-Ardenne); 
the D\'epartement des Ardennes;
and the Communaut\'e de Communes Ardenne Rives de Meuse.
We acknowledge the support of 
the CEA, CNRS/IN2P3, the computer centre CC-IN2P3
and 
LabEx UnivEarthS in France;
the Max Planck Gesellschaft, 
the Deutsche Forschungsgemeinschaft DFG, 
the Transregional Collaborative Research Center TR27, 
the excellence cluster ``Origin and Structure of the Universe''
and 
the Maier-Leibnitz-Laboratorium Garching in Germany;
the Ministry of Education, Culture, Sports, Science and Technology of Japan (MEXT) 
and
the Japan Society for the Promotion of Science (JSPS) in Japan;
the Ministerio de Econom\'ia, Industria y Competitividad (SEIDI-MINECO) under grants FPA2016-77347-C2-1-P and MdM-2015-0509 in Spain;
the Department of Energy and the National Science Foundation;
the Russian Academy of Science, the Kurchatov Institute and 
the Russian Foundation for Basic Research (RFBR) in Russia;
the Brazilian Ministry of Science, Technology and Innovation (MCTI), 
the Financiadora de Estudos e Projetos (FINEP), 
the Conselho Nacional de Desenvolvimento Cient\'ifico e Tecnol\'ogico (CNPq), 
the S\~ao Paulo Research Foundation (FAPESP)
and the Brazilian Network for High Energy Physics (RENAFAE) in Brazil.
}

\clearpage

\section*{\label{sec:methods}Appendix} 

In this section, we shall provide a complementary description about the following topics:
a) neutrino oscillation mechanism,
b) the DC detectors,
c) the TnC IBD selection and BG rejection strategy,
d) the cross-check and validation of the \ang\ measurement, 
e) the shape-only characterisation of the spectral distortion 
and
f) the mean cross-section per fission measurement.

\subsection*{Neutrino Oscillation}
Neutrino oscillation changes the neutrino flavour periodically while it travels in space-time. 
Its basic mechanism is the same as the Pauli equation extended to three basic states. 
The oscillation probability, shown in Eq.~\ref{eq:Intro:App_Probability}, is the same as the spin precession in a magnetic field.
The neutrino states $\nu_e$ and $\nu_\mu$ correspond to the spin-up and spin-down states. 
$\theta$ corresponds to the polar angle of the applied magnetic field and the $\Delta m^2/4E$ term corresponds to the angular velocity of the precession slowed down due to the time dilation of the relativistic effect. 
There is no parameter in the oscillation form that corresponds to azimuthal angle of the magnetic field, $\phi$, which is included in the Pauli equation as an imaginary phase.
However, actually there are at least 3 neutrinos and 
for three-state oscillations, an imaginary phase of the transition amplitudes manifests itself as a physical effect, which may violate CP symmetry.

\subsection*{\label{sec:detector}The Double Chooz Near \& Far Detectors}

\noindent
The DC detectors are identically designed and are expected to provide identical responses after calibration. This was a key requirement to achieve cancellation of most detection and energy systematics when measuring \ang\ with both detectors.
During the first phase of the experiment with only the far detector taking data (FD-I with no ND), the DC main effort was devoted to control of the SD systematics thus yielding unprecedented rate \& shape uncertainties.
This precision has led to among the most precise SD \ang\ and $\langle \sigma_f\rangle$ measurements to date.
A key element for a high SD physics performance is the detector simulation accuracy. This is given by a detailed description of the optical interface, material and geometry, based on Geant4~\cite{Ref_G4},
and the electronics response including 
PMT (photo-multiplier tube)
electronics signals and sampling.
Both data and simulation followed identical calibration methods, full event reconstruction and energy response definition.
In this way, the simulation is treated as an independent detector able to yield an accurate representation of the data needed for SD systematics reduction.

The DC cylindrical detector design consists of several internal volumes and an external muon detector.
From inside out, the main detector (labelled {\it Inner-Detector} or ID) is subdivided into three optically coupled volumes: 
a {\it Neutrino-Target} (GdT: 10.3~m$^3$ of liquid scintillator loaded with 1g/l Gd), 
a {\it Gamma-Catcher} (GC: 22.6~m$^3$ of liquid scintillator with no loading), 
and 
a {\it Buffer} (100~m$^3$ of non-scintillating oil).
The ID is fully surrounded by the {\it Inner Veto} (IV) detector.
Both ID and IV are topped by the {\it Outer Veto} (OV) muon detector.
The IV is a $\sim$0.5~m thick liquid scintillator detector except for a small fraction in the chimney region for the 4$\pi$ tagging of cosmic $\mu$'s.
However, the IV has been able to tag also external rock $\gamma$'s (i.e. as an anti-Compton veto) and cosmic neutrons caused by cosmic $\mu$'s going through the nearby rock.
The OV is a tracking plastic scintillator for cosmic $\mu$'s with a few cm positioning resolution~\cite{Ref_DCRecOV}.
Both the ID and IV share one common architecture, i.e.~a detector chimney (centre-top) allowing access to the sensitive volumes mainly for calibration deployment.
The chimney is an acceptance hole to the IV (only a small fraction of the top), thus allowing for stopped $\mu$'s to reach the ID undetected by the IV.
An extension of the OV was placed to cover much -- but not all -- of the IV where it was missing acceptance due to the chimney.

The detector readout employs PMTs and an 8-bit 
Flash-ADC
electronics sampling at 500~MHz~\cite{Ref_FADC}.
The ID and IV are instrumented with 390 10"~\cite{Ref_DC10PMT} and 78 8"
PMTs, respectively operated at a nominal gain of 10$^7$.
Custom-made front-end electronics ensure the pulse dynamics, including pre-amplification (gain $\sim$10) and match the Flash-ADC specification for accurate sampling, thus minimising digitisation artefacts.
A dedicated global self-trigger system~\cite{Ref_DCTrigger} was used to identify IBD interactions using a combined and tuneable energy and multiplicity criterion.
The readout energy threshold was kept low ($\sim$0.3~MeV) thus having negligible impact on the relevant physics which starts well above 0.5~MeV.
After each trigger, the deadtime-less DAQ causes 256~ns of Flash-ADC sampling to be read out, thus allowing for \textgreater150~ns of light scintillating pulse sampling.
The Flash-ADC waveforms are reconstructed offline to infer time and charge information per channel yielding \textless1~ns time resolution and efficient pulse identification from signals with less than 1/5 of a photo-electron (PE).
This information seeds the subsequent high-level event reconstruction stages such as vertex position, pulse shape, energy, etc.
Further details on event reconstruction were covered in~\cite{Ref_Gd-III}.
The Flash-ADC information, once reconstructed, can also provide unique pulse-shape event classification~\cite{Ref_DCOPS,Ref_DCOMEGA}.
A stringent event-by-event background rejection was possible with negligible detection systematics. 

The detector configuration was slightly modified for the FD-II running by implementing a few DAQ optimisations identified during the FD-I period as well as by the increase of the gain per channel by $\sim$2$\times$ to reduce the effect of the Flash-ADC-induced non-linearity~\cite{Ref_FADC_NL}.
Upon the discovery of the spontaneous light emission~\cite{Ref_DCLN} effect in the FD, the ND PMTs were covered with black films.
The ND exhibits barely no light noise effect, as opposed to the FD.
However, careful analysis of the FD data has demonstrated that almost full rejection was possible.
Indeed, the PMT light noise has demonstrated to have negligible effect to IBD selection, as also corroborated with the ND data with negligible effect.
Hence, both FD-II and ND have all PMTs switched on\footnote{During the commissioning of the FD-I 14 PMT's with the highest spontaneous light emission were left off since the experimental rejection of those events have not been fully proved.} to maximise response linearity and uniformity across both detectors.
So, the ND configuration is identical to that of FD-II.

\begin{figure*}[ht!]
	\centering
	\includegraphics[scale=0.45]{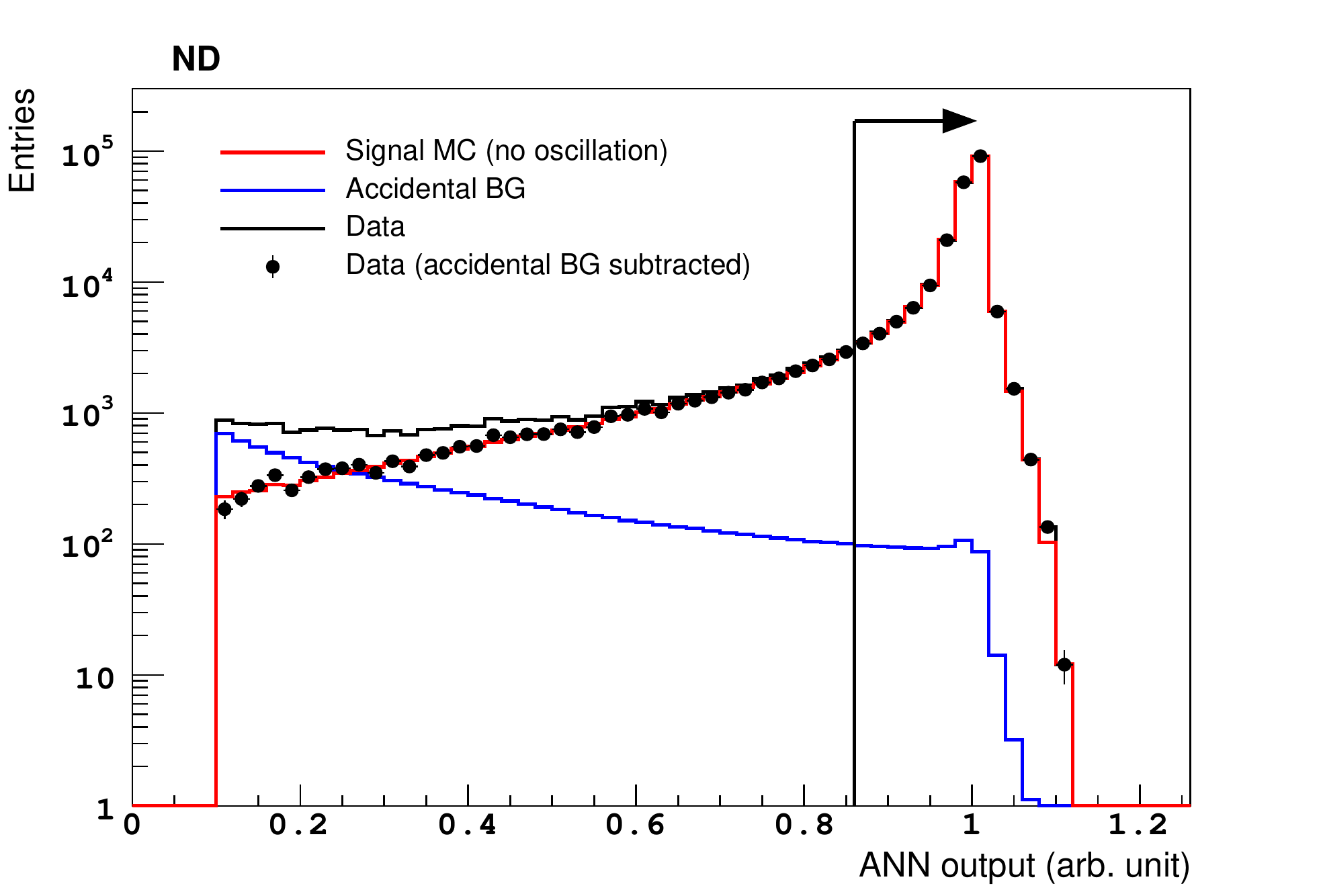}
	\includegraphics[scale=0.45]{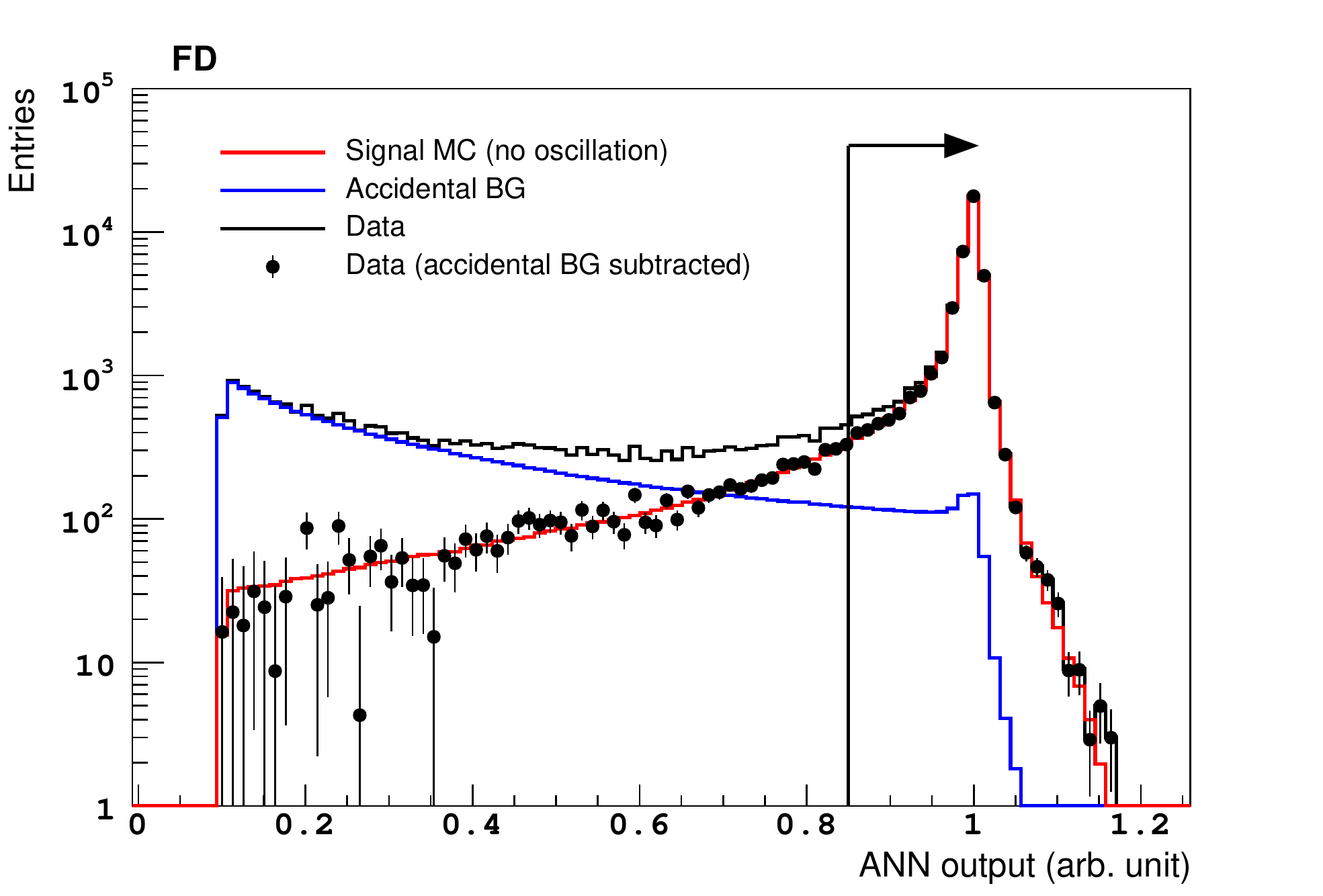}
	\caption{\small 
	{\bf TnC Selection Artificial Neural Network Definition.}
	The ND (left) and FD (right) Artificial Neural Network (ANN) cut definition are shown.
	Each plot shows full data (black-solid) and accidental BG only (blue-solid) curves.
	The remaining data upon BG subtraction is shown (black-points) represents correlated events, which are signal IBD-like.
	The IBD MC (solid red), with no BG's, is contrasted against the data.
	Sizeable differences between the FD and the ND ANN output are dominated by the different signal to BG contamination of each detector.
	The ND has $\sim$10$\times$ better signal to accidental BG.
	The FD has lower statistics.
	The MC exhibits excellent agreement to data across the entire dynamic for both detectors.
	A similar ANN definition had been demonstrated for FD-I data~\cite{Ref_H-III}.
	The ANN per detector cut was optimised to reduce the FD BG and to match a slight prompt spectral distorsion in both detectors (not shown explicitly).
	The latter is key to ensure an unbiased rate+shape \ang\ measurement.
	Such a distorsion is known to arise from the $\Delta$r$^{\mbox{\tiny prompt-delay}}$ variable slightly dependent on the prompt energy.
	Hence, the indicated ANN cut are slightly different for ND (0.86) and FD (0.85).
	This causes a 1.3\% difference in rate normalisation, corroborated with data to a few per mille precision.
	}
	\label{Fig-ANN}
\end{figure*}

The main energy estimator relies on the PE sum detected by all PMT, each estimated from charge integration converted into PE by a PMT gain calibration. Digitisation effects at low charge are corrected. The energy scale is defined by the 2.22~MeV H-n peak where all detectors and MC are equalised in response. The detector response non-uniformity and stability are corrected using the H-n gamma from spallation neutron captures.
For both detectors, data and simulation responses are independently calibrated using the same method.
The full volume uniformity (data and MC) and stability (data) systematics for each detector were both estimated to be less than 0.5\%.
The relative difference between the ND and FD energy responses were identical to \textless0.1\% between [0.5,10.0]~MeV, using the prompt spectrum with the same $^{252}$Cf source in both detectors.
The systematic uncertainties associated with the detector energy response were evaluated to have a small impact on the signal normalisation ($\leq$0.01\%).
Uncertainties on the energy linearity model are evaluated by source calibration data and further constrained in the \ang\ fit by the shape information.

A small amount of Gd, consistent with a mean concentration of $(1.1\pm0.4)\micro$g/l, was found in the GC of the ND.
Fast-neutrons allow for tomographic maps of both H-n and Gd-n captures for leak positioning and time evolution characterisation.
All observables compromised by the leak have been replaced in favour of leak-robust quantities, including the TnC detection technique strategy used for IBD detection, unlike the past n-Gd and n-H selections.
There is no trace of any leak in the FD.
The proton number uncertainty (NT: 0.3\%, GC: 1.1\%) is the dominant contributor to the signal normalisation uncertainty. The error on the GC proton number relies on all measurements available to date, and will be re-assessed with dedicated measurements during the detector dismantling.
To calculate the proton-number, the absolute liquid mass and the relative H fraction in the scintillator need to be known. 
Whereas masses can be measured at the 0.1\% level and below, the H fraction can be determined at the 1\% level with standard technologies such as CHN elemental analysis. 
The Gd-scintillator was produced in one batch and is chemically identical for both detectors. 
Therefore the uncertainty on the mass ratio can be estimated purely from the weight measurements including temperature effects. 
The ND to FD proton-number ratio for the Gd-scintillator is $1.0042\pm0.0010$. 
For the two GC volumes the masses are determined from the calculated volumes of the vessels and the measured liquid density.
 The dominant uncertainty for a SD is from the limited H fraction knowledge. 
However, a big fraction of this uncertainty is correlated among samples. 
The total ND to FD proton-number ratio in the GC including mass determinations and H fraction contributions is $1.0045\pm0.0067$.

\begin{table}[t]
\centering
\begin{tabular}{ L{3cm} | C{2.3cm} | C{2.7cm}}
{\bf Observable} & {\bf Condition} & {\bf Background}\\
\hline
{\bf $\mu$ Tagging} &&\\
Energy Deposited	& IV:\textgreater15MeV ID:\textgreater100MeV & through-$\mu$, stop-$\mu$ \\
\hline
{\bf Single} &&\\
Energy 		& $\geq$0.3~MeV 		& accidental\\
$\Delta$t($\mu$) 	& $\geq$1.25~ms 		& after $\mu$ activity\\
Light Noise veto	&  anomalous trigger 	& no ``light noise''\\
\hline
{\bf IBD Coincidence} &&\\
$\Delta$E$^{\mbox{\tiny prompt}}$ 		& [1.0,20.0]~MeV 	& keep IBD\\
ANN	Coincidence					& \textgreater0.85(FD)
\textgreater0.86(ND) 	& accidental\\
\multicolumn{1}{ r | }{$\Delta$E$^{\mbox{\tiny delay}}$\,~~~~~~~} 	& [1.3,10.0]~MeV 	& keep n-captures\\
\multicolumn{1}{ r | }{$\Delta$t$^{\mbox{\tiny prompt-delay}}$} 	& [0.5,800]~$\micro$s 	& accidental\\
\multicolumn{1}{ r | }{$\Delta$r$^{\mbox{\tiny prompt-delay}}$} 	& $\leq$1.2~m 	& accidental\\
$\Delta$t(unicity)$^{\mbox{\tiny prompt}}$				& [-800,900]~$\micro$s	& multi-coincidence\\
\hline
{\bf BG Vetoes}&&\\
IV$^{\mbox{\tiny prompt}}$	 	& IV activity 		& fast-n, stop-$\mu$\\
IV$^{\mbox{\tiny delay}}$	 		& IV activity 		& anti-Compton $\gamma$\\
OV$^{\mbox{\tiny prompt}}$	 	& OV activity 	& fast-n, stop-$\mu$\\
Stop-$\mu^{\mbox{\tiny delay}}$	& chimney  		& Michel-e$^\pm$ stop-$\mu$\\
Spallation$^{\mbox{\tiny prompt}}$	& n tagged $\mu$ 	& $^9$Li, $^{12}$B\\
\hline
\end{tabular}
\caption{\small	 
	{\bf TnC IBD Selection Criteria \& Background Rejection}.
	The complete TnC selection definition is here detailed, including selection criteria and BG vetoes. 
	The type of background rejected by each cut is also highlighted.
	}
	\label{Tab_Selection}
\end{table}

\begin{figure}[t]
	\centering
	\includegraphics[scale=0.47]{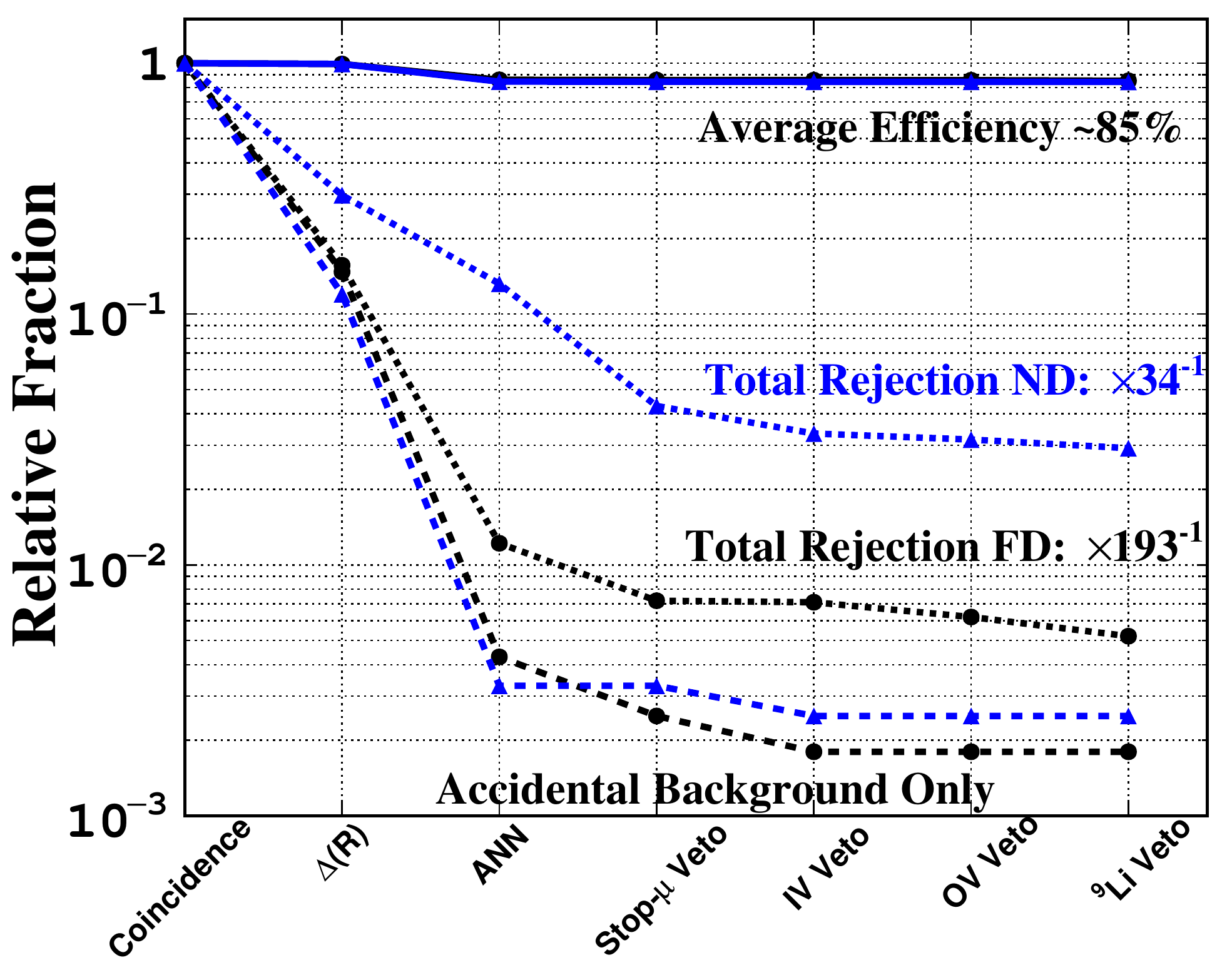}
	\caption{\small 
	{\bf TnC Efficiency \& Background Rejection.}
	The evolution of the TnC selection is illustrated in terms of IBD selection efficiency (solid lines), total BG rejection (dotted lines) and the accidental BG rejection (dashed lines).
	The estimation of the total BG rejection uses 17\,days of 0-reactor data.
	The average singles rate per detector is $\sim$10~s$^{-1}$. 
	The first criterion corresponds to a time of [0.5,800]$~\micro$s as a ``loose'' coincidence with a [1.0,20.0]~MeV prompt and the [1.3,10.0]~MeV delayed triggers.
The rates are 2291~day$^{-1}$ (FD) and 2375~day$^{-1}$ (ND), which imply a rejection factor of $\sim$375 relative to singles.
	These numbers provide an absolute scale to the all other shown below.
	The $\Delta$r$^{\mbox{\tiny prompt-delay}} \leq$1.2~m condition yields some important reduction. 
	However, major accidental BG rejection is only obtained by the ANN with a $\sim$400 rejection factor.
	After the ANN, the challenging correlated cosmogenic BG dominates the total BG rate, as expected due to the shallow overburden.
	The FD is better shielded.
	Extra rejection uses the cosmogenic vetoes.
	The overall rejection factors are $\sim$193 (FD) and $\sim$34 (ND) relative to the loose coincidence.
	}
	\label{Fig-Rejection}
\end{figure}

\subsection*{\label{sec:selection}The TnC Selection \& Background Rejection}

The main rationale behind the TnC selection is the wider aperture of the delayed energy window to accept neutron captures from all elements present in the detector; i.e. the H-C elements and the loaded Gd.
The TnC selection can be regarded as an effective combination of the well understood Gd-n and H-n selections~\cite{Ref_Gd-III,Ref_H-III}.
Table~\ref{Tab_Selection} summarises the TnC selection criteria categorised into $\mu$-tags, singles, IBD coincidences and BG vetoes.
The description follows.

The selection starts by identifying $\mu$'s using the $\mu$-tagging criterion such that the sample of singles excludes events just after $\mu$'s.
The after $\mu$'s events -- rich in cosmic neutron captures -- are not used for IBD selection.
However, those events are used for self-calibration and regular detector response and capturing monitoring across the entire volume and detector live time.

The IBD selection starts by imposing the IBD coincidence, in both time and space, between the prompt and the delayed neutron capture candidates.
The ANN imposes the multi-dimensional prompt-delayed correlation trained to reject random coincidences; i.e. most of the accidental BG.
Since the Gd sample has negligible accidental BG contribution (a few per months)~\cite{Ref_Gd-III}, the ANN impact the contribution mainly from the GC volume affecting mainly the H sample~\cite{Ref_H-III}.
The input variables are
$\Delta$t$^{\mbox{\tiny prompt-delay}}$,
$\Delta$r$^{\mbox{\tiny prompt-delay}}$
and
$\Delta$E$^{\mbox{\tiny delay}}$.
The first two variables exploit the fact that random coincidences exhibit, by definition, aleatory distributions.
For example, the $\Delta$t$^{\mbox{\tiny prompt-delay}}$ is flat for random coincidences as opposed to correlated events which exhibit the characteristic neutron-capture time distribution.
Instead, $\Delta$r$^{\mbox{\tiny prompt-delay}}$ grows with r$^3$ (saturating within the detector acceptance) for random coincidences, while correlated events are contained within about 1~m.
Thus, the ANN mainly exploits these very different measured patterns between correlated and random coincidences to select correlated events, such as IBD's.
The $\Delta$E$^{\mbox{\tiny delay}}$ has a minor impact since the TnC uses a wide range. 
There are few events below (\textless1.3~MeV) and hardly any event above (\textgreater10~MeV) the energy range considered.
The ANN, however, eliminates signal candidates in the range [3.0,3.5]~MeV, as illustrated in Fig.~\ref{Fig-TNC} where the signal to background ratio is expected to be low.
The ANN training benefits from copious and clean samples for both signal and accidental BG.
The ANN is expected to be immune to correlated coincidences, such as the IBDs and most cosmogenic BGs, specially the $^9$Li BG\footnote{The ANN leads to relatively small rejection fraction of fast-neutrons via the spatial coincidence condition since those events could extend over a somewhat larger volume.}.
The signal relies on the IBD MC demonstrating an excellent agreement with data in both input and output variables over the full dynamics considered.
The ANN optimisation criteria discussion is illustrated in Fig.~\ref{Fig-ANN}.
Here, the challenge is matching data to MC for both the $\Delta$r$^{\mbox{\tiny prompt-delay}}$ variable and energy resolution, both considered by DC in previous publications.
The BG relies on high statistics samples of accidental BG obtained using the standard ``off-time window'' technique, as described in~\cite{Ref_Gd-III}.
So, the overall training of the ANN makes the rejection insensitive to events with a neutron in the final state, which includes IBD's, all cosmogenic BG and some remaining accidental BG.
This is consistent with the observation, again in Fig.~\ref{Fig-TNC}, that the remaining accidental BG (dark grey) exhibits a clear irreducible H-n peak contribution.
An important intrinsic feature of the ANN is that it internally constructs the best combined selection from all the input variables to achieve an optimal signal identification.
This yields a superior background rejection as compared to the combination of simple individual cuts, as done in the Gd-only selection -- irrelevant then since the accidental BG rate was negligible.
The overall accidental BG rejection factor is $\geq$400$\times$ using the ANN, as illustrated in Fig.~\ref{Fig-Rejection}.
The input ANN selection is so widely open that the selection efficiency in the GdT is close to 100\% in the detector center due to the lower average capture time caused by the presence of Gd.
The ANN mainly affects events in the GC which are most sensitive to the rejection of external accidental BG.
Much of the ANN performance was explored, tuned and demonstrated during the H-n capture \ang\ measurement~\cite{Ref_H-III}.
Once the ANN-based IBD coincidence is defined, the unicity condition imposes that only two-fold coincidences are considered (consistent with IBD's).
After the ANN, cosmogenic BG dominates, especially in the ND due to the lower overburden.

The last stage is the application of vetoes targeting the cosmogenic BGs.
The vetoes exploit the fact that BG events deposit energy in different detector sensitive layers used to tag large fractions (typically about $\sim$50\%) of the cosmogenic BG.
Further details on the definition of all vetoes is found in~\cite{Ref_Gd-III,Ref_H-III} where they were first used and fully described.
The performance of the IBD TnC selection is illustrated in Fig.~\ref{Fig-Rejection}, including the overall BG rejection factors and selection efficiency for both the ND and FD.
The larger ND signal rate compensates for the larger BG, as compared to the FD.

\begin{figure}[t!]
	\centering
	\includegraphics[scale=0.48]{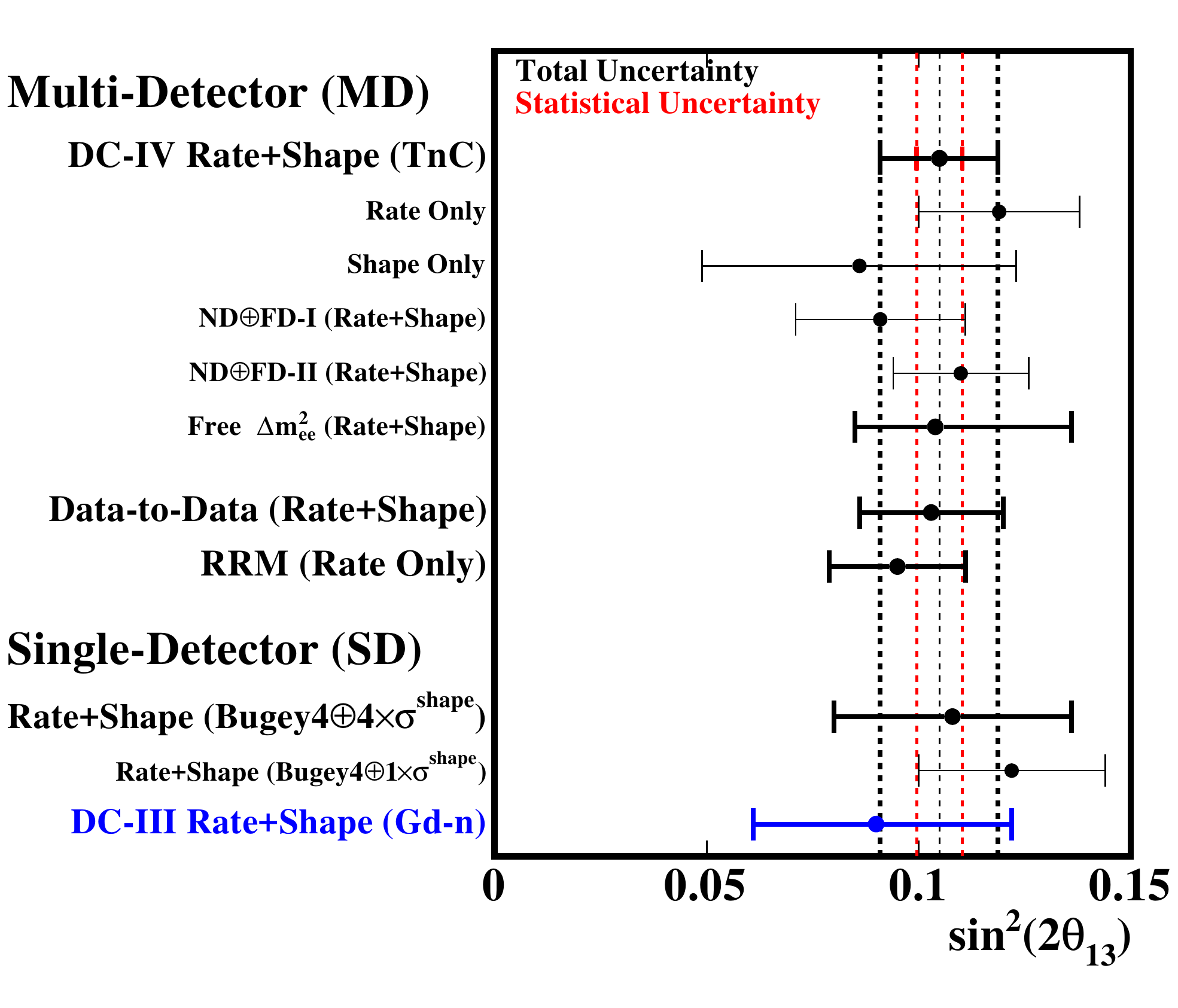}
	\caption{\small {\bf Scrutiny of the $\mathbf{\theta_{13}}$ Measurement.}
	The nominal \ang\ measurement (top) can be decomposed into 
	a) the rate-only and shape-only contributions,
	b) FD-I (no ND) and FD-II (iso-flux) contributions. 
	A measurement without marginalising over $|\Delta m^2_{ee}|$ as $(2.484\pm 0.036)\times10^{-3}$eV$^2$~\cite{Ref_NuFit3.1} is also shown.
	These numbers demonstrate that the nominal \ang\ measurement is dominated by the rate-only information (systematics limited) of the best FD-II iso-flux data sample.
	Furthermore, releasing the constraint on $|\Delta m^2_{ee}|$ does not impact the measured central value of \ang.
	Two alternative \ang\ measurements are also shown for comparison:
	a) the Data-to-Data
	and
	b) the Reactor Rate Modulation (RRM).
	Both are expected to be immune to the reactor model spectrum distortion while excellent agreement is found (details in text).
	Last, the FD-I+FD-II SD \ang\ measurements are also shown using two uncertainty prescriptions.
	The new data-driven prescription uses an increased 4$\sigma$ reactor model shape uncertainty.
	The standard reactor model prescription is also shown, indicating a bias on the result.
	The agreement of the SD and MD, with the more conservative uncertainty and the much better $\chi^2$/DoF, suggests the new prescription provides a better treatment of the data.
	The previous FD-I only SD \ang\ measurement~\cite{Ref_Gd-III} (blue) is shown for reference. 
	Bugey4 must be used in all SD to protect the rate normalisation.
	}
	\label{Fig-T13KST}
\end{figure}

\subsection*{\label{sec:t13cx}Further $\mathbf{\theta_{13}}$ Measurement Cross-Checks}

As part of the internal validation of the nominal \ang\ measurement, we have investigated several fits.
Fig.~\ref{Fig-T13KST} summarises the most relevant results.
Most of them have already been mentioned in past sections when demonstrating the minimal dependence of the \ang\ measurement to the observed spectral distortion between the data and the prediction model. However, we shall highlight two other \ang\ measurements with lower dependence on the model.

First is the so called {\bf Data-to-Data} (D2D) rate+shape \ang\ fit. This fit determines \ang\ by the comparison of the observed FD spectra to the prediction extracted from the ND spectrum. The model calculation is used only as the ratio of near versus far spectra, reducing the sensitivity to potential common model biases.
The FD-I can be compared to the ND data with a correction using MC ratios, whose uncertainty is significantly smaller than the statistical precision.
Likewise, common (or correlated) inter-detector effects are expected to fully cancel.
The value obtained is \sang=$0.103\pm0.017$ (sensitivity 0.0164 and $\chi^2$/DoF: 28/37).
The output inter-detector ratio exhibits no traces of spectral mismatch and looks identical to Fig.~\ref{Fig-Ratio}-(left).
The excellent agreement found to the main result (deviation $\delta \leq 0.002$ units) further supports the negligible impact of the reactor model in the MD case.
Second is the so called {\bf Reactor Rate Modulation} (RRM) rate-only \ang\ fit that has been articulated in past publications for the SD only~\cite{Ref_DCRRM}.
The rate-only implementation makes the RRM spectral distortion independent, by construction, even though the reactor rate prediction is indeed used.
This time the RRM prompt energy window has been confined to the [1.0,8.5]~MeV range with a slightly better IBD signal to BG.
This way, the [8.5,12.0]~MeV window data provide an independent cosmogenic BG counter (no shape information) used in the fit for a higher precision BG constraint.
The RRM fit remains very sensitive to the BG input, unlike the nominal rate+shape implementation reassessing the BG using the spectral shapes.
Despite this limitation, the novel RRM method yields a more precise BG with similar precision to the nominal \ang\ measurement.
The best value obtained is \sang=$0.095\pm0.016$ ($\chi^2$/DoF = 12/14).
Overall consistency is found within uncertainties.

\begin{figure*}[t]
	\centering
	\includegraphics[scale=0.75]{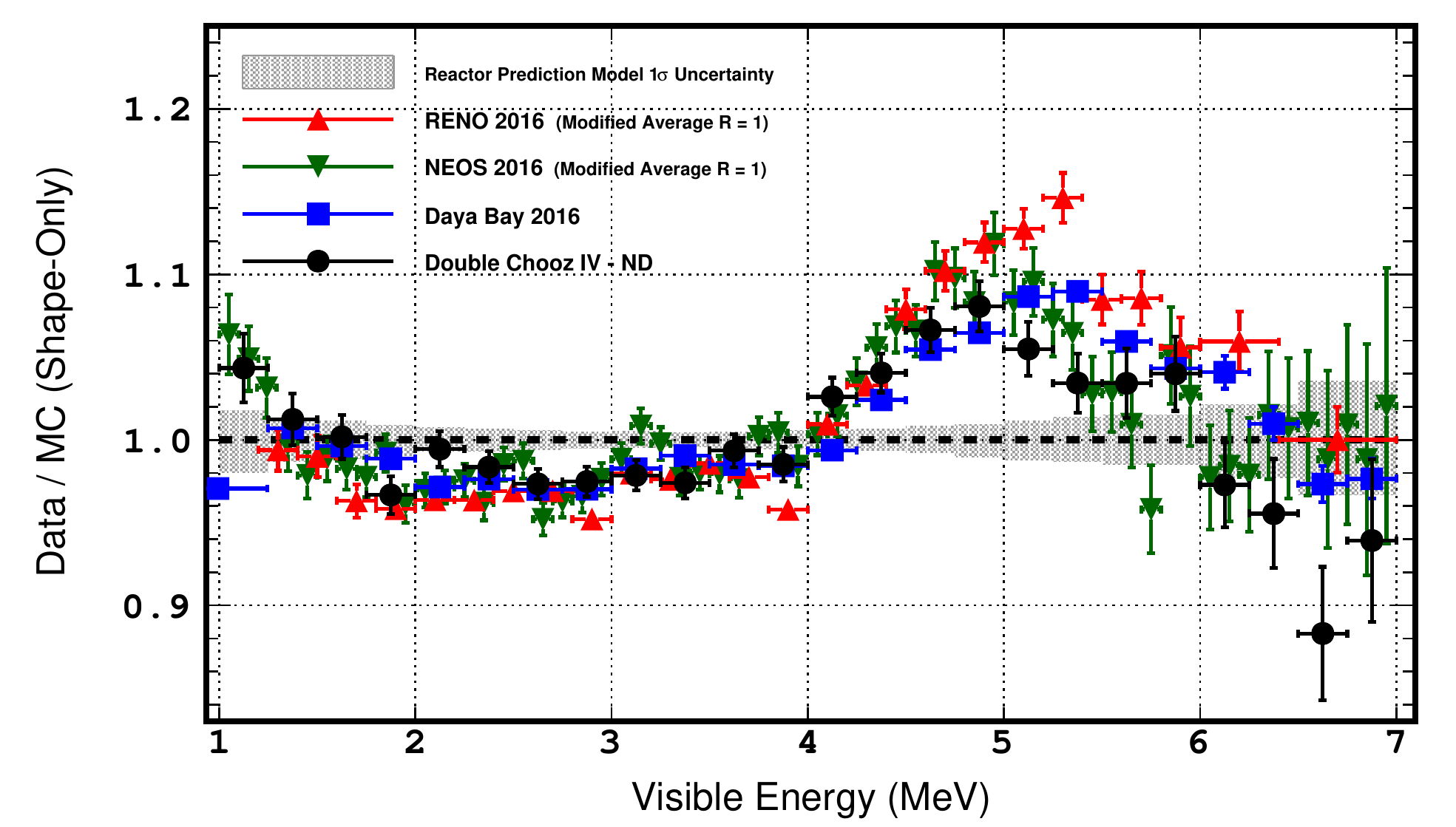}
	\caption{\small 
	{\bf Shape-Only Reactor Spectral Distortion.}
	The data to prediction spectral ratio for the latest DC-ND (black), DYB~\cite{Ref_DYB-MCSpF} (blue), RENO~\cite{Ref_RENO} (red), NEOS~\cite{Ref_NEOS} (green) are shown, exhibiting a common dominant pattern predominantly characterised by the 5~MeV excess.
	Small differences across experiments are still possible but unresolved so far.
	The Bugey3~\cite{Ref_B3} (not shown) is the only experiment known not to reproduce this structure.
	This remains an issue.
	The RENO and NEOS normalisation has been modified relative to publications to ensure the shape-only condition (average R\,=\,1) is met.
	The reactor model prediction shape-only uncertainty is shown in grey, which is significantly smaller than the dominant rate-only uncertainties.
	Since the same reactor model prediction is used, this uncertainty is expected to remain a representative guideline to all experiments.
	The 5~MeV excess is compensated by a deficit region [1.5,4.0]~MeV for all experiments due to the shape-only condition.
	A good agreement is found between DC and Daya Baya data throughout the entire energy range.
	The non-trivial match among different experiments suggests that most detector and part of the reactor effects are accurately reproduced by the MC, thus cancelling across in R. 
	This implies that the common reactor prediction model inaccuracies are expected to dominate the observed distortion.
	This is consistent with the fact that all other experiments use the same prediction strategy.
	}
	\label{Fig-SpecDistorsion}
\end{figure*}

\subsection*{\label{sec:distortion}The Reactor Model Structure}

Since 2011, there has been much debate about the rate deficit emerging with the revision of the reactor flux prediction~\cite{Ref_Huber,Ref_Moller}, including possible new particle physics~\cite{Ref_2010MCSpF,Ref_SterileReview}. 
Before, the predicted shape appeared to be well reproduced by the world best data sample provided by the Bugey3 experiment~\cite{Ref_B3}. 
Today, the situation has changed by some new observations.
In the following, we focus on the precise characterisation of those features while their specific origin and mechanism remains unknown. 
We exploit the shape-only basis even though this basis might not fully describe the data by ignoring the rate information. 
In fact, rate effects are large and there could be rate-to-shape correlations. 
Regardless, the shape-only basis allows for some simplified comparison among experiments and factorises out any rate normalisation effect. 
A rate+shape analysis was addressed in Fig.~\ref{Fig-Ratio}.

Upon the oscillation fit (including corrections for \ang, BG, normalisation, etc.) the ND number of events data to MC ratios (R$^N$) are R$^N$(ND)\,=\,$0.943\pm0.022$ and R$^N$(ND$\oplus$Bugey4)\,=\,$0.995\pm0.019$, respectively, without and with the Bugey4 constraint.
The R$^N$(ND) exhibits a 2.6$\sigma$ rate deficit, where the uncertainty is largely dominated by the prediction normalisation of the reactor model, as illustrated in Fig.~\ref{Fig-MCSpF}.
The quoted significance could be lower in case the reactor model uncertainty would be underestimated. 
The R(Bugey4) illustrates the excellent agreement found between Bugey4 and DC data. 
Most experiments are consistent with these observations.

The significant observation of a spectral distortion by the reactor-\ang\ experiments, dominated by an excess of events around  5~MeV, has shed new light on the prediction capability of the reactor antineutrino spectrum. Its existence suggests the presence of sizeable shape inaccuracies in addition to the aforementioned rate-only deficit. 
The 5~MeV excess was first reported~\cite{Ref_LALtalk} and published by DC in 2014~\cite{Ref_Gd-III}. 
Confirmations by RENO and DYB were reported shortly after. 
Today, there is a significant DC ND data to model disagreement when considering the full energy range.
The ND and FD spectra, once corrected for \ang, match within $\sim$1$\sigma$(stat) averaged over the full spectra.
This is expected as effective cancellation across detectors has been demonstrated in Fig.~\ref{Fig-Ratio}, as a part of the \ang\ measurement consistency. In addition, this distortion scales with reactor power: 6.4$\sigma$ (ND) and 7.1$\sigma$ (FD). 
So, an unknown new BG hypothesis is ruled out. 
The reactor model prediction and/or a residual non-linearity in the detector energy response~\cite{Ref_EnergySaclay} are both possible hypotheses. 
More exotic hypotheses have also been suggested~\cite{Ref_HuberLine}.

Further insight on the origin of the spectral distortion is gained when comparing the data to prediction ratios provided by the DC, DYB, RENO and NEOS~\cite{Ref_NEOS} experiments, as summarised in Fig.~\ref{Fig-SpecDistorsion}. 
From the point of view of shape distortions, the leading order pattern is well reproduced by all latest experiments. 
Other past experiments are not conclusive but CHOOZ and Goesgen~\cite{Ref_Goesgen} data show a similar pattern.
However, the Bugey3 experiment (not shown) exhibits a rather flat and featureless ratio spectrum, thus inconsistent with the shown experiments. 
The data to prediction ratio is expected to cancel (or suppress) the following contributions: 
a) the known and calibrated energy non-linearities, 
b) the detection effects (selection, vetoes, background subtraction bias), 
c) the $\theta_{13}$ spectral distortion, 
d) the reactor fuel burn-up 
and 
e) the overall shape of the reactor prediction, common to all experiments. 
However, energy response could be more complex since different detectors might have different capabilities to resolve spectral features. 
Nevertheless, most detector specific systematics are expected to be suppressed as a consequence of the MC tuning campaign to match the calibration data. 
So, while detector effects cannot be a priori ruled out, they are expected to be sub-dominant and not a priori identical across different experimental setups. 
Instead, the external reactor model, common to all experiments here considered, cannot be tuned to yield a data to prediction match.

A good agreement has been found between DC ND and the DYB ND's.
Their similar energy resolutions allow a simpler direct comparison. 
While all experiments plotted in Fig.~\ref{Fig-SpecDistorsion} show similar behaviour, the RENO data seem to hint at a larger 5~MeV excess. 
NEOS, located in the same RENO reactor plant, also appears consistent with DC data. 
By re-binning the DC data to the DYB binning, the compatibility of both data sets can be studied. 
The DC data has been corrected to account for the non-linearity constrained during the \ang\ fit and \ang\ itself.
The DC uncertainties consider statistics and systematics on both BG and energy. 
DC and DYB data are found consistent within uncertainties over the entire IBD energy spectrum. 
The best agreement manifests below 6~MeV. 
This observation disfavours detector driven effects as the main cause of the spectral distortion. 
Hence, the dominant cause appears to be the prediction inaccuracies in the reactor model.
More data should further resolve the issue.

Future efforts should scrutinise the data to prediction ratio in the rate+shape basis for completeness, as the shape-only basis remains incomplete. 
So far, most experimental data appear consistent with a reactor model origin for the observed spectral distortion as the leading order effect. 
Sub-dominant detector and/or reactor effects might still exist, possibly explaining some of the residual differences across experiments. 
So, today the undistorted Bugey3 spectrum remains an issue to be understood to be able to yield a coherent experimental vision. 
The need for further nuclear physics effort to improve the reactor model predictions remains a critical topic for high precision reactor neutrino physics. 
This is particularly important for any SD experiments. 
Last, a debate remains whether the main discrepancy feature is only the 5~MeV excess region since that part exhibits R$\to$1, as illustrated in Fig.~\ref{Fig-Ratio}, in a rate+shape treatment.
A preliminary more conservative reactor model error budget remains a pending issue with important impact to past and present experiments, so that their data and uncertainties are properly treated.

\subsection*{The Mean Cross-Section per Fission}

The mean cross-section per fission, or $\langle \sigma_f\rangle$, is defined as:

\begin{equation}
\langle \sigma_f\rangle = \frac{N(\bar\nu_e) }{N_{p} \times \epsilon} \times {(\sum_{r=B1,B2} \frac{\langle P_{th}\rangle_r}{4\pi L_r^2 \times  \langle E_f\rangle_r})}^{-1}~{\mbox{[cm$^2$/fission]}} 
\nonumber
\end{equation}

\noindent
where $\langle \sigma_f\rangle$ provides a reactor neutrino interaction probability measured over the integrated reactor spectrum.
$N(\bar\nu_e)$ stands for the measured neutrino rate (after BG substraction and correction for the \ang\ driven oscillation) in the considered detector,
$N_{p}$ is the proton-number in the target volume,
$\epsilon$ is the average absolute detection efficiency,
$\langle P_{th}\rangle_r$ is the average thermal power per reactor,
$L_r$ is the baseline per reactor
and
$\langle E_f\rangle_r$ is the average energy per fission per reactor~\cite{Ref_EperF}.
The main physical value and meaning of the $\langle \sigma_f\rangle$ is that it provides a measure of the total reactor neutrino integrated flux measurement per reactor which can be used as a common reference across experiments.
By correcting $\langle \sigma_f\rangle$ for the IBD cross-section a flux could be obtained (i.e.~neutrino/fission).

\begin{table}[t]
\centering
\begin{tabular}{ L{3.2cm} | C{2.1cm} }
{\bf Uncertainty (\%)} & {\bf ND}\\
\hline
Proton Number      	& 0.66\\
Thermal Power     		& 0.47\\
TnC Selection      		& 0.24\\
Background		     	& 0.18\\
Energy per Fission  		& 0.16\\
\ang\ Correction     		& 0.16\\
Statistics     			& 0.22\\
\hline
Total 				& 0.97\\
\end{tabular}
\caption{\small
	{\bf $\mathbf{\langle \sigma_f\rangle}$ Uncertainty Breakdown}.
	With a total uncertainty of about 1\%, the mean cross section per fission measured with the near detector is the most precise measurement to date. 
	The total uncertainty is dominated by the uncertainty on the proton number and on the reactor thermal power. 
	The detection related systematics, including proton-number, could still improve in DC while the reactor one is expected to be irreducible.
	\label{tab:mcspf}
	}
\end{table}

DC provides the most precise $\langle \sigma_f\rangle$ thus superseding the Bugey4 measurement, whose precision is 1.4\%.
Both measurements are compatible with one another as illustrated in Fig.~\ref{Fig-MCSpF}.
The systematics breakdown of the ND is summarised in Table~\ref{tab:mcspf} yielding just below 1\% precision for the first time.
Other experiments might use the DC $\langle \sigma_f\rangle$ in the same way DC has been using the $\langle \sigma_f\rangle$ of Bugey4; i.e.~to factorise out the rate (or normalisation) bias of the reactor prediction model. 
As demonstrated in the \ang\ measurement discussion, $\langle \sigma_f\rangle$ continues to be a critical value for past and future SD reactor experiments.
The $\langle \sigma_f\rangle$ measured by experiments using commercial nuclear reactors can slightly differ because of the different fuel composition. Differences typically at the 0.5\% level are expected when several reactor cycles are considered. 
For comparison with other experiments, two pieces of information are necessary.
The ND average fission fractions
52.0\%~($^{235}$U), 8.7\%~($^{238}$U), 33.3\%~($^{239}$Pu) and 6.0\%~($^{241}$Pu).
And, also the $\langle \sigma_f\rangle$ per isotope
6.69~($^{235}$U), 10.10~($^{238}$U), 4.36~($^{239}$Pu) and 6.05~($^{241}$Pu)
in units of $\times10^{-43}$cm$^{2}$ per fission, as obtained from~\cite{Ref_DYB-MCSPPI}.

The FD $\langle \sigma_f\rangle$ is, by definition, consistent with the ND upon \ang\ correction.
The FD alone yields a precision close 1.1\%.
Further precision improvements are expected in the future for $\langle \sigma_f\rangle$, especially if the dominant proton precision is improved upon detector dismantling.


\end{document}